\documentclass[10pt,journal,letterpaper]{IEEEtran}

\pdfoutput=1 

\usepackage{graphicx}
\usepackage{verbatim}
\usepackage{url}
\usepackage{amsmath}
\usepackage{amsfonts}
\usepackage{amssymb}
\usepackage[latin1]{inputenc}
\usepackage{delarray}
\usepackage{subfigure}
\usepackage{setspace}
\usepackage{paralist}
\usepackage{enumerate}

\usepackage{hyperref}
\IEEEoverridecommandlockouts

\newtheorem{lemma}{Lemma}[section]
\newtheorem{proposition}{Proposition}[section]

\newtheorem{theorem}{Theorem}[section]

\newtheorem{remarks}{Remarks}[section]
\newtheorem{corollary}{Corollary}[section]

\newcommand{\remove}[1]{}

\begin{document}
\title{Optimal Forwarding in Delay Tolerant Networks with Multiple Destinations}
\author{Chandramani~Singh,~\IEEEmembership{Student Member,~IEEE,}
Eitan~Altman,~\IEEEmembership{Fellow,~IEEE,}
Anurag~Kumar,~\IEEEmembership{Fellow,~IEEE,}\\
and~Rajesh~Sundaresan,~\IEEEmembership{Senior Member,~IEEE}
\thanks{This is an extended version of a paper that appeared in WiOpt~2011.}
\thanks{This work was supported by
the Indo-French Centre for the Promotion of Advanced
Research~(IFCPAR) Project 4000-IT-1, by DAWN~(an Associates
program of INRIA, France), and by the Department of Science 
and Technology, Government of India.}
\thanks{Chandramani~Singh, Anurag~Kumar and Rajesh~Sundaresan are with the Department of Electrical Communication Engineering
Indian Institute of Science Bangalore, India~(email: \{chandra,~anurag,~rajeshs\}@ece.iisc.ernet.in). Eitan~Altman is with INRIA, Sophia-Antipolis, France~(email: Eitan.Altman@sophia.inria.fr). }
}

\maketitle

\begin{abstract}

We study the trade-off between delivery delay and energy consumption
in a delay tolerant network in which a message~(or a file) has to be delivered to each of several
destinations by epidemic relaying. In addition to the destinations, there are
several other nodes in the network that can assist in relaying the message.
We first assume that, at every instant,
all the nodes know the number of relays carrying the
packet and the number of destinations that have received the packet. We formulate the problem
as a controlled continuous time Markov chain and derive the optimal closed loop control~(i.e., forwarding policy).
However, in practice, the intermittent connectivity in the network implies that the nodes may not have the required perfect knowledge
of the system state.
To address this issue, we obtain an ODE~(i.e., a deterministic fluid) approximation for the
optimally controlled Markov chain.
This fluid approximation also yields an asymptotically optimal open loop policy.
Finally, we evaluate the performance of the deterministic policy over finite networks.
Numerical results show that this policy performs close to the optimal closed loop policy.
\end{abstract}

\section{Introduction}

{\it Delay tolerant networks}~(DTNs)~\cite{comnet-dtn.fall03dtn-architecture}
 are sparse wireless ad hoc networks with highly mobile nodes.
In these networks, the link between any two nodes is up when these are
within each other's transmission range, and is down otherwise. In
particular, at any given time, it is unlikely that there is a complete route between a source and its
destination.

We consider a DTN in which a short message~(also referred to as a
{\it packet}) needs to be delivered to multiple~(say $M$)
destinations. There are also $N$ potential relays that do not
themselves ``want'' the message but can assist in relaying it to the
nodes that do. At time $t=0$, $N_0$ of the relays have copies of the
packet. All nodes are assumed to be mobile. In such a network, a
common technique to improve packet delivery delay is {\it epidemic}
relaying~\cite{comnet-dtn.vahdat-becker00epidemic-routing}. We
consider a controlled relaying scheme that works as follows.
Whenever a node~(relay or destination) carrying the packet meets a
relay that does not have a copy of the packet, then the former has
the option of either copying or not copying. When a node that has
the packet meets a destination that does not, the packet can be
delivered.

We want to minimize the delay until a significant fraction (say $\alpha$) of the destinations receive the packet;
we refer to this duration as {\it delivery delay}.
Evidently, delivery delay can be reduced if the number of carriers of
the packet is increased by copying it to relays. Such copying can not be
done indiscriminately, however, as every act of copying between two
nodes incurs a transmission cost. Thus, we focus
on the problem of the control of packet forwarding.

\noindent
{\bf Related work:} Analysis and control of DTNs with a single-source
and a single-destination has been widely studied. Groenevelt et
al.~\cite{comnet-dtn.groenevelt-etal05message-delay-mobile-networks}
modeled epidemic relaying and two-hop relaying using Markov chains.
They derived the average delay and the number of copies generated until
the time of delivery. Zhang et
al.~\cite{comnet-dtn.zhang-etal07epidemic-routing} developed a unified
framework based on ordinary differential equations~(ODEs) to study epidemic
routing and its variants.

Neglia and Zhang~\cite{ctrltheory-dtn.neglia-zhang06optimal-delay-power-tradeoff}
were the first to study the optimal control of relaying in
DTNs with a single destination and multiple relays. They
assumed that all the nodes have perfect knowledge of the number of nodes carrying the packet.
Their optimal closed loop control is a threshold policy - when a relay that does not have a copy
of the packet is met, the packet
is copied if and only if the number of relays carrying the packet is below a threshold.
Due to the assumption of complete knowledge, the reported performance is a lower bound for the cost in a real
system.

Altman et al.~\cite{stochctrl-dtn.altmanetal10monotone-forwarding-policies} addressed
the optimal relaying problem for a class of {\it monotone relay strategies} which
includes epidemic relaying and two-hop relaying. In particular,
they derived {\it static} and {\it dynamic} relaying policies.
Altman et al.~\cite{ctrltheory-dtn.altman-etal09decentralized-stochastic-control}
considered optimal discrete-time two-hop relaying. They also employed stochastic
approximation to facilitate online estimation of network parameters.
In another paper, Altman et al.~\cite{ctrltheory-dtn.altman-etal10optimal-activation-transmission-control} considered
a scenario where active nodes in the network continuously spend energy
while {\it beaconing}. Their paper studied the joint problem of node activation
and transmission power control. These works~(\cite{stochctrl-dtn.altmanetal10monotone-forwarding-policies,ctrltheory-dtn.altman-etal09decentralized-stochastic-control,ctrltheory-dtn.altman-etal10optimal-activation-transmission-control})
heuristically obtain fluid approximations for DTNs and study open
loop controls. Li et al.~\cite{stochctrl-dtn.lietal10optimal-opportunistic-forwarding}
considered several families
of open loop controls and obtain optimal controls within each family.

Deterministic fluid models expressed as ordinary differential equations have been used to approximate
large Markovian systems. Kurtz~\cite{stochproc.kurtz70limits-markov-processes} obtained sufficient
conditions for the convergence of Markov chains to such fluid limits. Darling~\cite{stochproc.darling02fluid-limits} and subsequently, Darling and Norris~\cite{stochproc.darling-norris08differential-equation-approximations} generalized Kurtz's results. Darling~\cite{stochproc.darling02fluid-limits} considers the scenario when the  Markovian system satisfies the conditions in~\cite{stochproc.kurtz70limits-markov-processes} only over a subset. He shows that the scaled processes converge to a fluid limit until they exit from this subset. Darling and Norris~\cite{stochproc.darling-norris08differential-equation-approximations} generalize the conditions for convergence, e.g., uniform convergence of the mean drifts of Markov chains and Lipschitz continuity of the limiting drift function, prescribed in~\cite{stochproc.kurtz70limits-markov-processes}.  Gast and Gaujal~\cite{stochctrl.gast-gaujal10mean-field-nonsmooth} address the scenario where the limiting drift functions are not Lipschitz continuous. They prove that under mild conditions, the stochastic system converges to the solution of a differential inclusion. Gast et al.~\cite{stochctrl.gast-etal10mean-field-MDPs} study an optimization problem on a large Markovian system. They show that solving the limiting deterministic problem yields an asymptotically optimal policy for the original problem.

\noindent
{\bf Our Contributions:} We formulate the problem as a controlled
continuous time Markov
chain~(CTMC)~\cite{stochctrl.bertsekas07dpoc-vol2}, and obtain the
optimal policy~(Section~\ref{forwarding-policy}). The optimal policy
relies on complete knowledge of the network state at every node,
but availability of  such information is constrained by the same connectivity
problem that limits packet delivery. In the incomplete information
setting, the decisions of the nodes would have to depend upon their beliefs
about the network state. The nodes would need to update their
beliefs continuously with time, and also after each meeting with
another node. Such belief updates would involve maintaining a complex
information structure and are often impractical for nodes with limited memory and
computation capability. Moreover, designing closed loop controls based on beliefs is a difficult
task~\cite{stochctrl-dtn.singhetal10dtn-twohop}, even more so in our
context with multiple decision makers and all of them equipped with
distinct partial information.

In view of the above difficulties, we adopt the following approach.
We show that when the number of
nodes is large, the optimally controlled network evolution is well
approximated by a deterministic dynamical system~(Section~\ref{asym-opt-forward}).
The existing differential equation approximation results for
Markovian systems~\cite{stochproc.kurtz70limits-markov-processes,stochproc.darling02fluid-limits}
do not directly apply, as, in the optimally controlled Markov chain that arises in our problem,
the mean drift rates are discontinuous and do not converge uniformly.
We extend the results to our problem setting in our Theorem~\ref{assym-optimality} in Section~\ref{asym-opt-forward}.
Note that the differential inclusion based approach of Gast and Gaujal~\cite{stochctrl.gast-gaujal10mean-field-nonsmooth}
is not directly applicable in our case, as it needs uniform
convergence of the mean drift rates.
The limiting deterministic dynamics then suggests a deterministic control
that is asymptotically optimal for the finite network problem, i.e.,
the cost incurred by the deterministic control approaches
the optimal cost as the network size grows.
We briefly consider the analogous control of two-hop forwarding~\cite{comnet-wireless.grossglauser-tse02mobility-adhoc-networks} 
in Section~\ref{sec:two-hop}.
Our numerical results
illustrate that the deterministic policy performs close to the
complete information optimal closed loop policy for a wide range of
parameter values~(Section~\ref{num-results}).

In a nutshell, the ODE approach is quite common in the modeling of such problems.
Its validity in situations without control is established by Kurtz~\cite{stochproc.kurtz70limits-markov-processes}, 
Darling and Norris~\cite{stochproc.darling-norris08differential-equation-approximations}, etc.
We aim in this paper at rigorously showing the validity of this limit under control in a few DTN problems.

\section{The System Model}
\label{sec:sys-model}
We consider a set of $K: = M+N$ mobile nodes. These include $M$
destinations and $N$ relays. At $t=0$, a packet is generated and
immediately copied to $N_0$ relays~(e.g., via a broadcast from an infrastructure
network). Alternatively, these $N_0$ nodes can be thought of
as source nodes.

\subsubsection{Mobility model}
We model the point process of the {\it meeting instants} between pairs of nodes as independent Poisson point processes, each with rate $\lambda$. Groenevelt et al.~\cite{comnet-dtn.groenevelt-etal05message-delay-mobile-networks} validate this model for a number of common mobility models~(random walker, random direction, random waypoint). In particular, they establish its accuracy under the assumptions of small communication range and sufficiently high speed of nodes.

\subsubsection{Communication model}
Two nodes may communicate only when they come within transmission
range of each other, i.e., at {\it meeting instants}.
The transmissions are assumed to be instantaneous. We assume that
that each transmission of the packet incurs unit energy expenditure
at the transmitter.

\subsubsection{Relaying model}
We assume that a controlled epidemic relay protocol is employed.

Throughout, we use the terminology relating to the spread of
infectious diseases. A node with a copy of the packet is said to be
{\it infected}. A node is said to be {\it susceptible} until it receives a copy
of the packet from another infected node. Thus at $t=0$,  $N_0$
nodes are infected while $M+N-N_0$ are susceptible.

\subsection{The Forwarding Problem}
\label{sec:forward-problem} The packet has to be disseminated to all
the $M$ destinations. However, the goal is to minimize the duration
until a fraction $\alpha$~($\alpha < 1$) of the
destinations receive the packet.

\remove{
 The goal is to deliver the packet to a fraction
$\alpha$~($\alpha < 1$) of the $M$ destinations within a short
duration. Thanks to the intermittent connectivity, nodes only have
beliefs about the number of infected destinations. Let
$\mathcal{T}_d$ be a time such that with high probability the
fraction of infected nodes at $\mathcal{T}_d$ is close to $\alpha$.
More precisely, if $m(t)$ is the number of infected destinations at
time $t$, then for any $\epsilon > 0$,
\begin{equation*}
\lim_{K \rightarrow \infty} \mathbb{P}\left(\left|\frac{m(\mathcal{T}_d)}{M} - \alpha\right| > \epsilon \right) = 0.\footnote{Both, the number of destinations and the number of relays,
are scaled together such that their ratio, $\frac{M}{N}$, remains fixed.}
\end{equation*}
We want to shorten $\mathcal{T}_d$. }

At each meeting epoch with a susceptible relay, an infected
node~(relay or destination) has to decide whether to copy the
packet to the susceptible relay or not. Copying the packet incurs
unit cost, but promotes early delivery of the packet to the
destinations.
We wish to find the trade-off between these costs by minimizing
\begin{equation}
\label{eqn-objective}
\mathbb{E}\{\mathcal{T}_d + \gamma \mathcal{E}_c\}
\end{equation}
where $\mathcal{T}_d$ is the time until which at least $M_{\alpha}
:= \lceil \alpha M \rceil$ destinations receive the packet,
$\mathcal{E}_c$ is the total energy consumed
in copying, and $\gamma$ is the parameter that relates
energy consumption cost to delay cost. Varying $\gamma$ helps studying
the trade-off between the delay and the energy costs.

\remove{
\begin{remarks}
In the presence of complete information, every node knows when
the desired fraction of destinations receive the packet.
Thus no copying is done to the remaining susceptible destinations. In the case
of partial information copying to the susceptible destinations is stopped
at $\mathcal{T}_d$. This ensures that the fraction of infected destinations is
close to $\alpha$.
\end{remarks}
}
\section{Optimal Epidemic Forwarding}
\label{forwarding-policy}
We derive the optimal forwarding policy under the assumption that, at any instant of time,
all the nodes have full information about the number of relays carrying the
packet and the number of destinations that have received the packet.
This assumption will be relaxed in the next section.

\subsection{The MDP Formulation}
\label{sec:mdp-formulation}
Let $t_k, k = 1,2,\dots$ denote the meeting epochs of the infected
nodes~(relays or destinations) with the susceptible nodes. Let $t_0 :=
0$ and define $\delta_k := t_k - t_{k-1}$ for $k \geq 1$.

Let $m(t)$ and $n(t)$ be the numbers of infected destinations and
relays, respectively, at time $t$. In particular, $m(0) = 0$ and
$n(0) = N_0$, and the forwarding process stops at time $t$ if $m(t)
= M$. We use $m_k$ and $n_k$ to mean $M(t_{k}-)$ and $N(t_{k}-)$ which are the numbers
of infected destinations and relays, respectively,  just before the meeting epoch $t_k$.
Let $e_k$ describe the type of the susceptible node
that an infected node meets at $t_k$; $e_k \in \mathcal{E} :=
\{d,r\}$ where $d$ and $r$ stand for destination and relay,
respectively. The state of the system at a meeting epoch $t_k$ is
given by the tuple
\begin{equation*}
s_k := (m_k, n_k, e_k).
\end{equation*}
Since the forwarding process stops at time $t$ if $m(t) = M$, the
state space is $[M-1] \times [N_0:N] \times
 \mathcal{E}$.\footnote{We use notation $[a] = \{0,1,\dots,a\}$ and $[a:b] = \{a,a+1,\dots,b\}$ for
$b \geq a+1$ and $a,b \in \mathbb{Z}_+$.}

\begin{figure}[t]{
\centering
\includegraphics[height=1.6in]{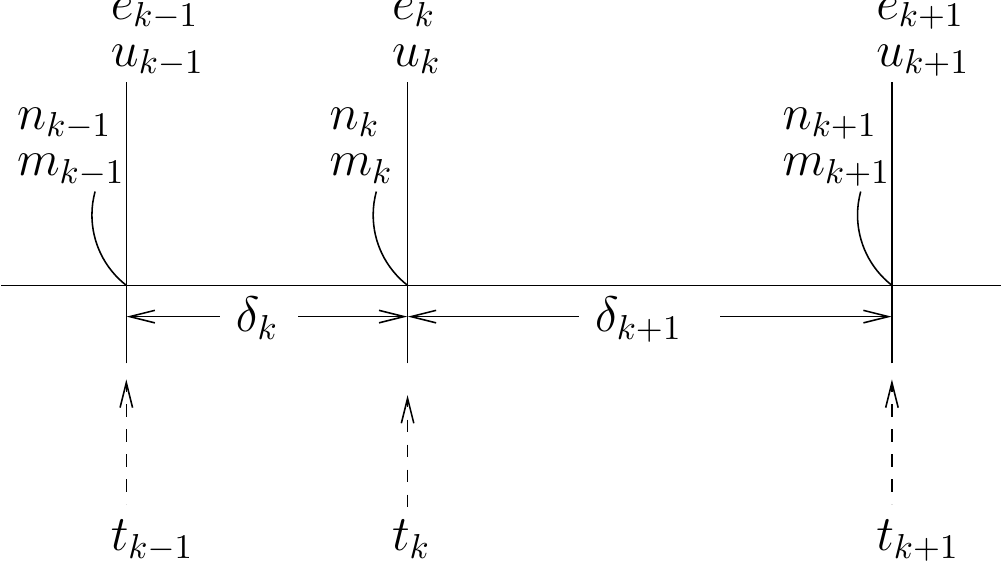}
\caption{\label{embedding}Evolution of the controlled Markov chain
$\{s_k\}$. Note that $(m_k,n_k)$ is embedded at $t_k-$, i.e., just before the meeting epoch.}}
\end{figure}

Let $u_k$  be the action of the infected node at
meeting epoch $t_k, k = 1,2,\dots$.
The control space is
$\mathcal{U} \in \{0,1\}$, where $1$ is for {\it copy} and $0$ is for {\it do not copy}.
The embedding convention described
above is shown in Figure~\ref{embedding}.

We treat the tuple $(\delta_{k+1},e_{k+1})$ as the random
disturbance at epoch $t_k$. Note that for $k = 1,2,\dots$, the time
between successive decision epochs, $\delta_k$, is independent and
exponentially distributed with parameter
$(m_k+n_k)(M+N-m_k-n_k)\lambda$. Furthermore, with ``w.p.'' standing for ``with probability'', we have
\begin{equation*}
e_k = \begin{cases}
                 d  & \mbox{w.p. } p_{m_k,n_k}(d) := \frac{M-m_k}{M+N-m_k-n_k}, \\
                 r  & \mbox{w.p. } p_{m_k,n_k}(r) := \frac{N-n_k}{M+N-m_k-n_k}.\end{cases}
\end{equation*}
\subsubsection{Transition structure}
From the description of the system model, the state at time $k+1$ is given by  $s_{k+1} = (m_k + u_k,
n_k, e_{k+1})$ if $e_k  = d$, and $s_{k+1} = (m_k, n_k + u_k,
e_{k+1})$ if $e_k  = r$. Recall that $e_{k+1}$ is a component in the
random disturbance. Thus the next state is  a function of the
current state, the current action and the current disturbance as
required for an MDP .

\subsubsection{Cost Structure}
For a state-action pair $(s_k,u_k)$
the expected single stage cost is given by
\begin{equation*}
g(s_k,u_k) = \gamma u_k + \mathbb{E}\left\{\delta_{k+1}1_{\{m_{k+1}
< M_{\alpha}\}}\right\},
\end{equation*}
where the expectation is taken with respect to the random
disturbance $(\delta_{k+1},e_{k+1})$. It can be observed that
\begin{align*}
g(s_k,u_k) = \begin{cases}
             \gamma u_k \mbox{ if }  s_k \mbox{ is such that } m_k \geq M_{\alpha} \\
                 \gamma \mbox{ if } s_k  = (M_{\alpha}-1,n,d) \mbox{ and } u_k = 1\\
                 \gamma u_k + C_d(s_k,u_k) \mbox{ otherwise},\end{cases}
\end{align*}
where
\[C_d(s_k,u_k) = \frac{1}{(m_k+n_k+u_k)(M+N-m_k-n_k-u_k)\lambda}\]
 is the mean time
until the next decision epoch. The quantity $\gamma$ is expended whenever $u_k = 1$, i.e., the action is to copy.

\subsubsection{Policies} A policy $\pi$ is a sequence of mappings $\{u^{\pi}_k, k = 0,1,2,\dots\}$,
where $u^{\pi}_k: [M-1] \times [N_0:N] \times \mathcal{E}
\rightarrow \mathcal{U}$. The cost of an admissible policy $\pi$ for
an initial state $s = (m,n,e)$ is
\begin{equation*}
J_{\pi}(s) = \sum_{k = 0}^{\infty}
\mathbb{E}\Big\{g(s_k,u^{\pi}_k(s_k))\big| s_0 = s\Big\}.
\end{equation*}
Let $\Pi$  be the set of all admissible policies. Then the optimal cost function is defined as

\begin{equation*}
J(s)  = \min_{\pi \in \Pi}J_{\pi}(s).
\end{equation*}
A policy $\pi$ is called stationary if $u^{\pi}_k$ are identical, say $u$, for all $k$.
For brevity we refer to such a policy as the stationary policy $u$.
A stationary policy $u^{\ast} \equiv \{u^{\ast},u^{\ast},\dots\}$ is optimal if $J_{u^{\ast}}(s) = J(s)$ for all states $s$.

\subsubsection{Total Cost}
We now translate the optimal cost-to-go from the first meeting instant into optimal total cost.
Recall that at the first decision instant $t_1$, the state
$s_1$ is $(0,N_0,r)$ or $(0,N_0,d)$ depending on whether the susceptible node that is met is a relay
or a destination.
The objective function~\eqref{eqn-objective} can then be restated as
\begin{align}
\mathbb{E}_{\pi}\{\mathcal{T}_d + \gamma\mathcal{E}_c\} =&~\frac{1}{\lambda N_0(M+N-N_0)} + \left(\frac{N-N_0}{M+N-N_0} \right. \nonumber\\
J_{\pi}(0,N_0,&r) + \left.\frac{M}{M+N-N_0}J_{\pi}(0,N_0,d)\right) \label{eqn:cost},
\end{align}
where the subscript $\pi$ shows dependence on the underlying policy. In the right hand side, the first term $\frac{1}{\lambda N_0(M+N-N_0)}$ is the average delay until the first decision instant which has to be borne under any policy.

\subsection{Optimal Policy}
\label{sec:opt-policy} Since the cost function $g(\cdot)$ is
nonnegative, Proposition~1.1 in~\cite[Chapter~3]{stochctrl.bertsekas07dpoc-vol2}
implies that the optimal cost function will satisfy the following
Bellman equation. For $s = (m,n,e)$,
\begin{align*}
J(s) &= \min_{u \in \{0,1\}} A(s,u) \\
\mbox{where } A(s,u) &= g(s,u) + \mathbb{E}\left(J(s')|s,u\right).
\end{align*}
Here $s'$ denotes the next state which depends on $s,u$ and the random disturbance
in accordance with the
transition structure described above. The expectation is taken with
respect to  the random disturbance. Furthermore, since the action
space is finite, there exists a stationary optimal policy $u^{\ast}$
such that, for all $s$, $u^{\ast}(s)$ attains minimum in the above
Bellman equation
~(see~\cite[Chapter~3]{stochctrl.bertsekas07dpoc-vol2}). In the
following we characterize this stationary optimal policy.

First, observe that it is always optimal to copy to a destination, that is, the optimal policy satisfies $u^{\ast}(m,n,d)
= 1$ for all $(m,n) \in [M-1] \times [N_0:N]$. Moreover, once a fraction $\alpha$ of the destinations have obtained the packet, no further delay cost is incurred, and so further copying to relays does not help: $u^{\ast}(m,n,r)
= 0$ for all $(m,n) \in [M_{\alpha}:M-1] \times [N_0:N]$.

Next, focus on a reduced state space $[M_{\alpha}-1] \times
[N_0:N] \times \{r\}$.
 \remove{ We cast the problem as a {\it stopping
problem}~\cite[Section~3.4]{stochctrl.bertsekas07dpoc-vol2}. Towards
this, we define an special $stop$ action which implies that no
copying is done to the susceptible relay met at present or those to
be met in the future. Alternatively, $stop$ is equivalent to a sequence
of actions $0,0,0,\dots$.
Clearly, if we choose
action $stop$ at a state $(m,n,r)$, the stopping cost, $sc(m,n,r)$,
is the expected time to go until the desired fraction of
destinations are infected. It can be easily verified that, for all
$(m,n) \in [M_{\alpha}-1] \times [N_0:N]$,
\begin{equation*}
sc(m,n,r) = (M-m)\gamma + \sum_{j =
m}^{M_{\alpha}-1}\frac{1}{\lambda(n+j)(M-j)}.
\end{equation*}
The one-step stopping set is defined as
\begin{equation*}
\mathcal{SS} = \left\{s \in [M_{\alpha}-1] \times [N_0:N] \times
\{r\} | sc(s) \leq min_{u \in \mathcal{U}}\left(g(s,u) +
\mathbb{E}(sc(s'|s,u,e'))\right) \right\},
\end{equation*}
where, as before, $e'$ and $s'$ denote the random disturbance and
the next state, respectively.
}
Consider the following {\it one step look
ahead policy}~\cite[Section~3.4]{stochctrl.bertsekas07dpoc-vol2}. At
a meeting with a susceptible relay, say when the state is $(m,n,r)$, compare
the following two action sequences.
\begin{enumerate}
\item $0s$: {\it stop}, i.e., do not copy to this relay or to any susceptible relays met in the future,
\item $1s$: copy to this relay and then {\it stop}.
\end{enumerate}
The costs to go corresponding to the action sequences
$0s$ and $1s$ are, respectively,
\begin{align*}
J_{0s}(m,n,r) &= (M-m)\gamma + \sum_{j = m}^{M_{\alpha}-1}\frac{1}{\lambda(n+j)(M-j)} \mbox{ and} \\
J_{1s}(m,n,r) &=  (M-m +1)\gamma + \sum_{j =
m}^{M_{\alpha}-1}\frac{1}{\lambda(n+j+1)(M-j)}.
\end{align*}
The {\it stopping set} $\mathcal{S_S}$ is defined to be
\begin{equation}
\mathcal{S_S} := \{(m,n,r):\Phi(m,n) \leq 0\} \label{eqn:stopping-set}
\end{equation}
where
\begin{align}
 \Phi(m,n) := &~J_{0s}(m,n,r) - J_{1s}(m,n,r)\nonumber \\
  = &~\sum_{j =m}^{M_{\alpha}-1}\frac{1}{\lambda(n+j)(n+j+1)(M-j)} - \gamma \label{eqn:Phi}
\end{align}
for all $(m,n) \in [M_{\alpha}-1] \times [N_0:N]$.
The one step look ahead policy is to copy to relay
when $(m,n,r) \notin \mathcal{S_S}$, and to stop copying otherwise.\footnote{We use the standard convention that a sum over an
empty index set is $0$. Thus $\Phi(m,n) = -\gamma$ if $m \geq
M_{\alpha}$. Consequently, for the states $[M_{\alpha}:M-1] \times [N_0:N] \times \{r\}$, one step-look ahead
policy prescribes {\it stop}. This is consistent with our earlier discussion.}

One step look ahead policies have been shown to be optimal for
stopping problems under certain
conditions~(see~\cite[Section~4.4]{stochctrl.bertsekas05dpoc-vol1}
and~\cite[Section~3.4]{stochctrl.bertsekas07dpoc-vol2}).
Let us reemphasize that our problem is not a stopping problem because an action $0$ now
is not equivalent to {\it stop} as the resulting state
is not a {\it terminal state};  a susceptible relay that is met in
the future may be copied even if the one met now is not.
However, we exploit the cost structure to prove that when an infected node meets a susceptible relay,
it can restrict attention to two actions: $1$~(i.e., copy now) and \emph{stop}~(i.e., do not
copy now and never copy again).
Subsequently, we also show that the above one step look ahead policy~(see~\eqref{eqn:stopping-set}) is optimal.
\begin{theorem}
\label{heu-optimality} The optimal policy $u^{\ast}:[M-1] \times
[N_0:N] \times \mathcal{E} \rightarrow \mathcal{U}$ satisfies
\begin{align*}
u^{\ast}(m,n,e) = \begin{cases}
 1, \mbox{ if } e = d,\\
 1, \mbox{ if } e = r \mbox{ and } \Phi(m,n) > 0, \\
\mbox{{\it stop}} \mbox{ if } e = r \mbox{ and } \Phi(m,n) \leq 0.\end{cases}
\end{align*}
\end{theorem}
\begin{IEEEproof}
Though the optimal policy is a simple stopping policy, the proof of its optimality is far from obvious. 
See Appendix~\ref{proof-heu-optimality}.
\end{IEEEproof}

\remove{
\begin{remarks}
\label{remark:Phi} Observe that $\Phi(m,n)$ is decreasing in $m$ for
a given $n$ and also decreasing in $n$ for a given $m$. Thus the
optimal policy has the following properties.\\
\begin{inparaenum}[1)]
\item If $u^{\ast}(m,n,r) = 0$, then $u^{\ast}(i,n,r) = 0$ for all $m < i < M$.\\
\item If $u^{\ast}(m,n,r) = 0$, then $u^{\ast}(m,j,r) = 0$ for all $n < j < N$.
\end{inparaenum}

Thus the optimal solution can be given a ``stopping''
interpretation. More precisely, if the packet is not copied at a
meeting with a susceptible relay, it is not copied to relays in
future meetings. A priori however, we did not know if such a
``stopping'' was optimal.
\end{remarks}
}
\remove{
\begin{remarks}
Also observe that
 \[
 J_{1s} (m,n,r) = J_{0s} (m,n+1,r) + \gamma
 \]
 Hence
\begin{align*}
A((m,n+1,r),) - \Phi(m,n) =
 J_{1s} (m,n,r) - J_{0s} (m,n,r)
 = J_{0s} (m,n+1,r) - J_{0s} (m,n,r)  + \gamma
 \end{align*}
 Since phi is decreasing in m, we conclude that
\begin{enumerate}
\item $ J_{0s} $ is convex in $m$, and thus so is $J_{1s}$
\item They are both submodular
\end{enumerate}
\end{remarks}
}

\begin{figure}[t]{
\centering
\includegraphics[height=2.5in]{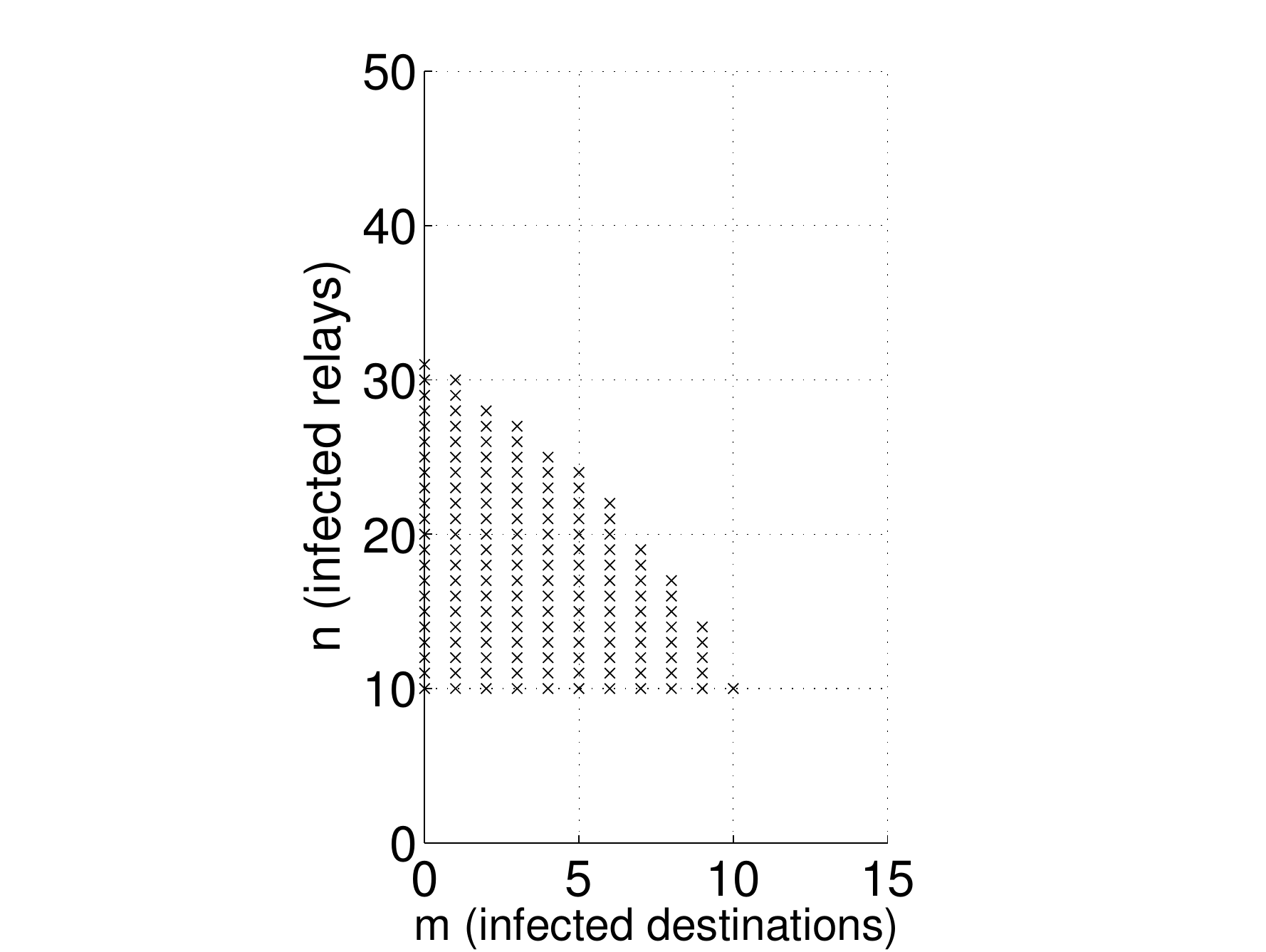}
\caption{An illustration of the optimal policy. The symbols 'X' mark
the states in which the optimal action~(at meeting with a relay) is
to copy}
 \label{fig:heu-policy}}
\end{figure}

We illustrate the optimal policy using an example. Let $M =15, N=
50, N_0 = 10, \alpha = 0.8, \lambda = 0.001$ and $\gamma = 1$. The
``$\times$'' in Figure~\ref{fig:heu-policy} are the
states where the optimal action~(at meeting with a relay) is to
copy. For example, if only $5$ destinations have the packet, then
relays are copied to if and only if there are $24$ or less infected
relays. If $7$ destinations already have the packet and there are
$19$ infected relays, then no further copying to relays is done.

\section{Asymptotically Optimal Epidemic Forwarding}
\label{asym-opt-forward}

In states $[M_{\alpha}-1] \times [N_0:N] \times \{r\}$, the optimal
action, which is governed by the function $\Phi(m,n)$, requires
perfect knowledge of the network state~$(m,n)$.
This may not be available to the decision maker due to intermittent
connectivity.
In this section, we derive an asymptotically optimal policy that
does not require knowledge of network's state but depends only on
the time elapsed since the generation of the packet. Such a policy
is implementable if the packet is time-stamped when generated and
the nodes' clocks are synchronized.

\subsection{Asymptotic Deterministic Dynamics}
\label{asym-det-dynamics}
 Our analysis closely follows
Darling~\cite{stochproc.darling02fluid-limits}. It is
straightforward to show that the equations that follow are the conditional expected
drift rates of the optimally controlled CTMC. For $(m(t),n(t)) \in
[M-1] \times [N_0:N]$, using the optimal policy in Theorem~\ref{heu-optimality}, we get
\begin{subequations}
\begin{align}
\frac{{\rm d}\mathbb{E}(m(t)|(m(t),n(t)))}{{\rm d}t} = &~\lambda (m(t) + n(t))(M-m(t)), \label{eqn:drift-1}\\
\frac{{\rm d}\mathbb{E}(n(t)|(m(t),n(t)))}{{\rm d}t} = &~\lambda (m(t) + n(t))(N-n(t)) \nonumber\\
& \ \ \ \ \ \ \ \ \ \ \ \ \ 1_{\{\Phi(m(t),n(t)) > 0\}}. \label{eqn:drift-2}
\end{align}
\end{subequations}

Recalling that $K = M+N$, the total number of nodes, we study large $K$ asymptotics. Towards
this, we consider a sequence of problems indexed by $K$.
The parameters of the $K$th problem are denoted using the superscript $K$.
Normalized versions of these parameters, and normalized versions of the system state
are denoted as follows:
\begin{equation}
\label{eqn:scalings}
\left. \begin{aligned} X = \frac{M^K}{K},~Y = \frac{N^K}{K},\\
X_{\alpha} = \frac{\alpha M^K}{K},~Y_0 = \frac{N_0^K}{K}, \\
\lambda^K = \frac{\Lambda}{K}, \gamma^K = \frac{\Gamma}{K},\\
~x^K(t) = \frac{m(t)}{K} \mbox{ and }y^K(t) = \frac{n(t)}{K}.
\end{aligned} \right\}
\end{equation}
\remove{
\begin{equation}
\label{eqn:scalings}
\left. \begin{aligned} X = \frac{M(K)}{K},~Y = \frac{N(K)}{K},\\
X_{\alpha} = \frac{\alpha M(K)}{K},~Y_0 = \frac{N_0(K)}{K}, \\
\lambda(K) = \frac{\Lambda}{K}, \gamma(K) = \frac{\Gamma}{K},\\
~x^K(t) = \frac{m^K(t)}{K} \mbox{ and }y^K(t) = \frac{n^K(t)}{K}.
\end{aligned} \right\}
\end{equation}
}
\begin{remarks}
The pairwise meeting rate and the copying cost must both scale down as $K$ increases.
Otherwise, the delivery delay will be negligible and the total transmission cost will be
enormous for any policy, and no meaningful analysis is possible.
\end{remarks}

For each $K$, we define scaled two-dimensional integer lattice
\begin{equation*}
\Delta^K = \left\{\left(\frac{i}{K} ,\frac{j}{K}\right): (i,j) \in [M^K-1] \times [N_0^K:N^K] \right\}.
\end{equation*}
$(x^K(t),y^K(t)) \in \Delta^K$. Also, for $(x^K(t),y^K(t)) \in \Delta^K$, using the notation in~\eqref{eqn:scalings}, the
drift rates in~\eqref{eqn:drift-1}-\eqref{eqn:drift-2}  can be rewritten as follows.
\remove{
\footnote{More precisely,
$(x^K(t),y^K(t))$ lies on a scaled two-dimensional integer lattice
of the from $(i/K ,j/K)$ for some $i,j \in \mathbb{Z}_+$.}
}
\begin{subequations}
\begin{align}
&\frac{{\rm d}\mathbb{E}(x^K(t)|(x^K(t),y^K(t)))}{{\rm d}t} \nonumber \\
& \ \ \ \ \ \ \ \ \ \ \ \ \ \ \ \ \ \ \ \ \ \ \ = f^K_1(x^K(t),y^K(t)) \nonumber \\
& \ \ \ \ \ \ \ \ \ \ \ \ \ \ \ \ \ \ \ \ \ \  := \Lambda(x^K(t)+ y^K(t))(X-x^K(t)), \label{eqn:f-K_1} \\
&\frac{{\rm d}\mathbb{E}(y^K(t)|(x^K(t),y^K(t)))}{{\rm d}t} \nonumber \\
& \ = f^K_2(x^K(t),y^K(t)) &  \nonumber \\
&:= \Lambda (x^K(t) + y^K(t))(Y-y^K(t))1_{\{\phi^K(x^K(t),y^K(t)) > 0\}}, \label{eqn:f-K_2}
\end{align}
\end{subequations}
where, for $(x,y) \in \Delta^K$,
\begin{equation}
\label{eqn:phi-K}
 \phi^K(x,y) := \sum_{j = Kx}^{\lceil KX_{\alpha} \rceil -1}\frac{1}{K\Lambda(y
+ \frac{j}{K})(y + \frac{j+1}{K})(X -\frac{j}{K})} - \Gamma.
\end{equation}
We also define $(x(t),y(t)) \in [0, X] \times [Y_0,Y]$ as functions
satisfying the following ODEs: $x(0)= 0, y(0) = Y_0$, and for $t
\geq 0$,
\begin{subequations}
\begin{align}
\frac{{\rm d}x(t)}{{\rm d}t} = &~f_1(x(t),y(t)) := \Lambda (x(t) + y(t))(X-x(t)),\label{eqn:f-1} \\
\frac{{\rm d}y(t)}{{\rm d}t} = &~f_2(x(t),y(t)) := \Lambda (x(t) + y(t))(Y-y(t)) \nonumber\\
                               & \ \ \ \ \ \ \ \ \ \ \ \ \ \ \ \ \ \ \ \ \ \ \ \ \ \ \ \ \ \ \ 1_{\{\phi(x(t),y(t)) > 0\}}\label{eqn:f-2}
\end{align}
\end{subequations}
where\footnote{We use the convention that an integral assumes the value $0$
if its lower limit exceeds the upper limit. So, $\phi(x,y) = -\Gamma$ if $x \geq X_{\alpha}$.}
\begin{equation}
\label{eqn:phi}
 \phi(x,y) = \int_{z = x}^{X_{\alpha}}\frac{{\rm d}z}{\Lambda(y + z)^2(X - z)} - \Gamma.
\end{equation}
Finally, we redefine the delivery delay $\mathcal{T}_d$~(see~\eqref{eqn-objective}) to be
\begin{align}
\tau^K &= \inf\{t \geq 0: x^K(t) \geq X_{\alpha}\},\label{eqn:tau-K} \\
\mbox{and }\tau &= \inf\{t \geq 0: x(t) \geq X_{\alpha}\}. \label{eqn:stop-dstns}
\end{align}
 Note that $\tau^K$ is a stopping time for the random
process $(x^K(t),y^K(t))$, whereas $\tau$ is a deterministic time instant. Since $f^K_1(x,y)$ is bounded away from
zero, $\tau^K < \infty$ with probability $1$. Similarly, on account of $f_1(x,y)$ being bounded away from
zero, $\tau < \infty$.

\begin{figure}[t]
\centering
\subfigure{\includegraphics[width=2.6in]{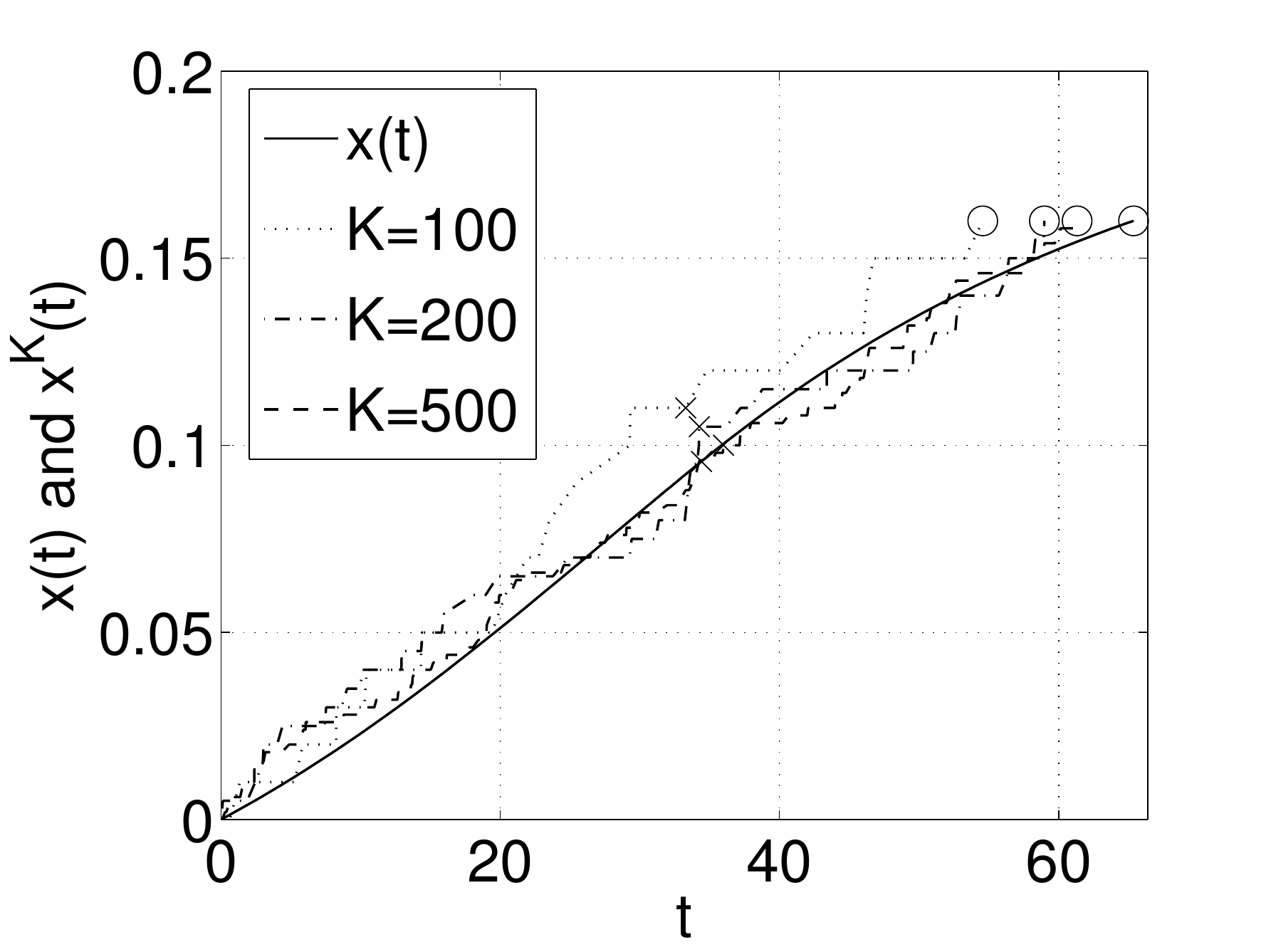}}
\subfigure{\includegraphics[width=2.6in]{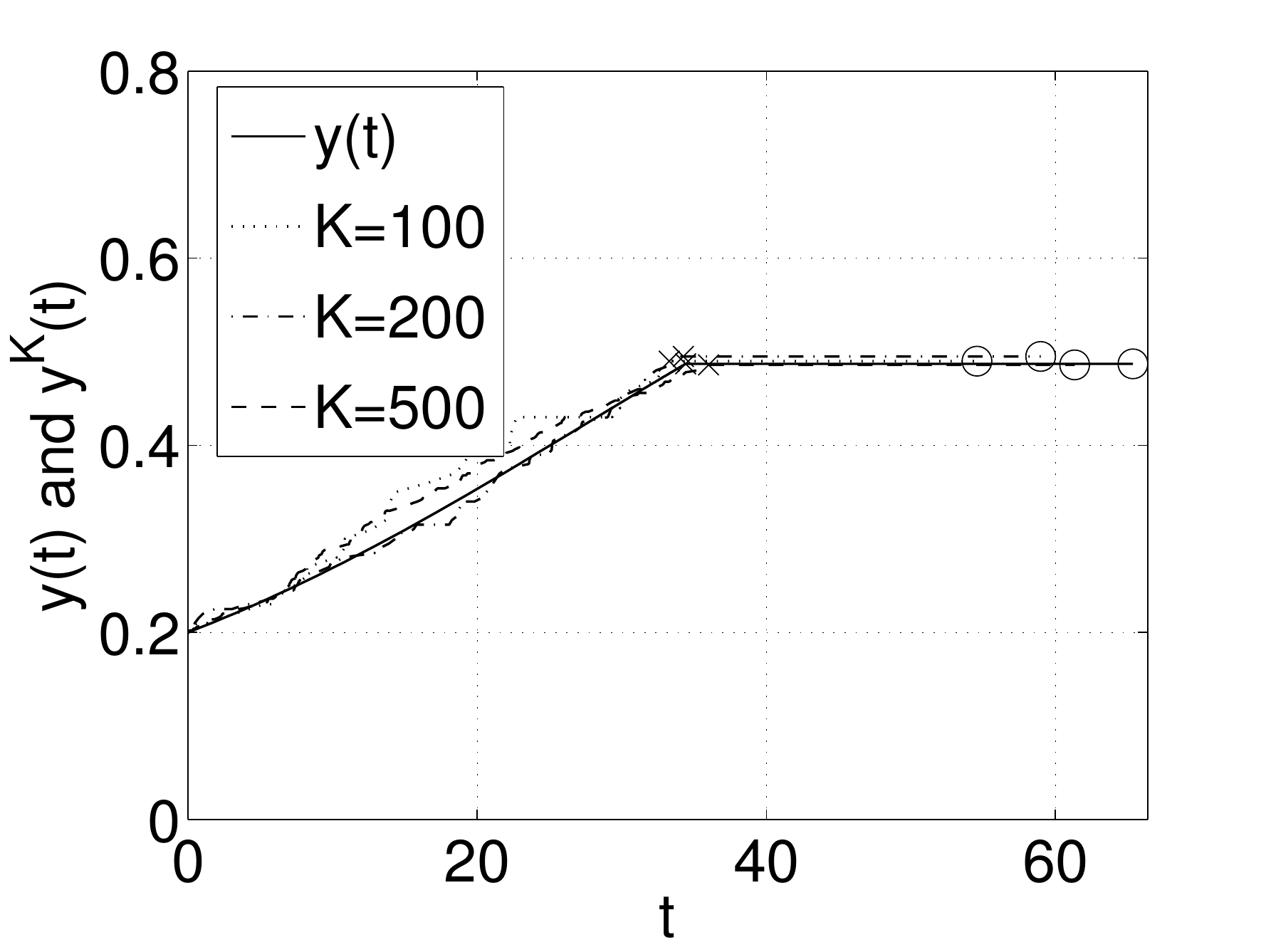}}
\caption{Simulation results: The top and bottom sub-plots
respectively show the fractions of infected destinations and relays
as a function of time. $(x^K(t),y^K(t))$ are obtained from a
simulation of the controlled CTMC, and $(x(t),y(t))$ from the ODEs.
The marker 'X' indicates the states at which copying to relays is
stopped whereas 'O' indicates the states at which a fraction $\alpha$
of destinations have the packet.} \label{fig:kurtz}
\end{figure}

Kurtz~\cite{stochproc.kurtz70limits-markov-processes} and Darling~\cite{stochproc.darling02fluid-limits}
studied convergence of CTMCs to the solutions of ODEs. The following are the hypotheses for
the version of the limit theorem that appears in Darling~\cite{stochproc.darling02fluid-limits}.
\begin{enumerate}[(i)]
\item $\lim_{K \rightarrow \infty} \mathbb{P}\left(\Vert(x^K(0),y^K(0) - (x(0),y(0))\Vert > \epsilon \right) = 0$;
\item In the scaled process $(x^K(t),y^K(t)$, the jump rates are $O(K)$ and drifts are $O(K^{-1})$;
\item $(f^K_1(x,y), f^K_2(x,y))$ converges to $(f_1(x,y), f_2(x,y))$ uniformly  in $(x,y)$;
\item $(f_1(x,y), f_2(x,y))$ is Lipschitz continuous.
\end{enumerate}
Observe that, in our case, only the first two hypotheses are satisfied.
In particular, $f^K_2(x,y)$ does not converge uniformly to $f_2(x,y)$,
and $f_2(x,y)$ is not Lipschitz over $[0, X_{\alpha}] \times [Y_0,
Y]$. Hence, the convergence results do not directly apply in
our context. Thankfully, there is some regularity we can exploit which
we now summarize as easily checkable facts.
\begin{enumerate}[(a)]
\item $\phi^K(x,y)$ converges uniformly to $\phi(x,y)$;
\item the drift rates $f_1(x,y)$ and $f_2(x,y)$ are bounded from below and above;
\item $f_1(x,y)$ is Lipschitz and $f_2(x,y)$ is locally Lipschitz; and
\item for all small enough $\nu \in \mathbb{R}$, and all $(x,y)$ on the graph of ``$\phi(x,y) = \nu$'', the direction in which the ODE
progresses, $(f_1(x,y),f_2(x,y))$, is not tangent to the graph.
\end{enumerate}
We then prove the following result which is identical to
~\cite[Theorem~2.8]{stochproc.darling02fluid-limits}.
\begin{theorem}
\label{assym-optimality}
Assume that $\alpha < 1$ and $Y_0 > 0$. Then, for every $\epsilon,\delta > 0$,
\begin{align*}
&\lim_{K \rightarrow \infty} \mathbb{P}\left(\sup_{0 \leq t \leq \tau}\Vert(x^K(t),y^K(t) - (x(t),y(t))\Vert > \epsilon \right) = 0, \\
&\lim_{K \rightarrow \infty} \mathbb{P}\left(|\tau^K - \tau| > \delta \right) = 0.
\end{align*}
\end{theorem}
\begin{IEEEproof}
See Appendix~\ref{proof-assym-optimality}.
\end{IEEEproof}
\remove{
Moreover, we also obtain the following result.
\begin{theorem}
\label{thm:asym-stop-time}
For every $\epsilon > 0$,
\begin{equation*}
\lim_{K \rightarrow \infty} \mathbb{P}\left(\left|\frac{m(\tau)}{M} - \alpha\right| > \epsilon \right) = 0.
\end{equation*}
\end{theorem}
\begin{IEEEproof}
\end{IEEEproof}
}
We illustrate Theorem~\ref{assym-optimality} using an example. Let $X = 0.2,Y = 0.8,\alpha = 0.8,Y_0 = 0.2,\Lambda = 0.05$
and $\Gamma = 50$. In Figure~\ref{fig:kurtz}, we plot $(x(t),y(t))$ and
sample trajectories of $(x^K(t),y^K(t))$ for $K = 100,200$
and $500$. We indicate the states at which the optimal policy stops copying to
relays, i.e., $\Phi^K(x^K(t),y^K(t))$ goes below $0$~(see Theorem~\ref{heu-optimality})
and the states at which the fraction of infected destinations crosses $X_{\alpha}$.
We also show the corresponding states in the fluid model.
The plots show that for large $K$, the fluid model captures the random dynamics
of the network very well.

\subsection{Asymptotically Optimal Policy}
Observe that $\phi(x,y)$ is decreasing in $x$ and $y$, both of which are nondecreasing with $t$. Consequently
$\phi(x(t),y(t))$ decreases with $t$. We define
\begin{equation}
\label{eqn:stop-relays}
\tau^{\ast} := \inf\{t \geq 0: \phi(x(t),y(t)) \leq 0\}.
\end{equation}
The limiting deterministic dynamics suggests the following policy
$u^{\infty}$ for the original forwarding problem.\footnote{Observe that the policy $u^{\infty}$ does not require
knowledge of $m$ and $n$. The infected node readily knows the type
of the susceptible node~($d$ or $r$) at the decision epoch.}

\begin{equation*}
u^{\infty}(m,n,e) = \begin{cases}
 1 \mbox{ if } e = d,\\
 1 \mbox{ if } e = r \mbox{ and } t \leq \tau^{\ast},\\
 0 \mbox{ if } e = r \mbox{ and } t > \tau^{\ast}.\end{cases}
\end{equation*}

We show that the policy $u^{\infty}$ is asymptotically optimal in the sense that
 its expected cost approaches the expected cost of the optimal
policy $u^{\ast}$ as the network grows.
Let us restate~\eqref{eqn:cost} as
\begin{align*}
\mathbb{E}^K_{\pi}\{\mathcal{T}_d + \gamma\mathcal{E}_c\} =&~\frac{1}{K\Lambda Y_0(1-Y_0)} + \left(\frac{Y-Y_0}{1-Y_0} \right. \\
&~\left.J_{\pi}(0,Y_0,r) + \frac{X}{1-Y_0}J_{\pi}(0,Y_0,d)\right) \label{eqn:cost}.
\end{align*}
We have used superscript $K$ to show the dependence of cost on the
network size. We then establish the following asymptotically optimality result.
\begin{theorem}
\label{theorem:asym-optimal}
\[
\lim_{K \rightarrow \infty} \mathbb{E}^K_{u^{\ast}}\{\mathcal{T}_d + \gamma\mathcal{E}_c\} =  \lim_{K \rightarrow \infty} \mathbb{E}^K_{u^{\infty}}\{\mathcal{T}_d + \gamma\mathcal{E}_c\} = \tau + \Gamma y(\tau^{\ast}).
\]
\remove{
\[
\lim_{K \rightarrow \infty} \left(\mathbb{E}^K_{u^{\infty}}\{\mathcal{T}_d + \gamma\mathcal{E}_c\} - \mathbb{E}^K_{u^{\ast}}\{\mathcal{T}_d + \gamma\mathcal{E}_c\}\right) = 0.
\]
}
\end{theorem}
\begin{IEEEproof}
See Appendix~\ref{proof-asym-optimal}.
\end{IEEEproof}
\begin{remarks}
\label{remark:cost-comparision}
Observe that we do not compare the limiting value of the optimal costs with the 
optimal cost on the~(limiting) deterministic system. In general, these two may differ.\footnote{In our case these two indeed match. See Appendix~\ref{hamiltonian} for a proof.} 
However, the deterministic policy $u^{\infty}$ can be applied on the finite $K$-node system. 
The content of the above theorem is that given any $\epsilon > 0$, cost of the policy $u^{\infty}$ 
is within $\epsilon$ of the optimal cost {\em on the $K$-node system} for all sufficiently large $K$.
\end{remarks}

\subsubsection*{Distributed Implementation}
The asymptotically optimal policy can be implemented
in a distributed fashion. Assume that all the nodes are time
synchronized.\footnote{In practice, due to variations in the clock
frequency, the clocks at different nodes will drift from each other.
But the time differences are negligible compared to the delays
caused by intermittent connectivity in the network. Moreover, when
an infected node meets a susceptible node, clock synchronization can
be performed before the packet is copied.} Suppose that the packet
is generated at the source at time $t_0$~(we assumed $t_0 = 0$ for
the purpose of analysis). Given the system parameters
$M,N,\alpha,N_0,\lambda$ and $\gamma$, the source first extracts
$X,Y,X_{\alpha},Y_0,\Lambda$ and $\Gamma$ as
in~\eqref{eqn:scalings}. Then, it calculates
$\tau^{\ast}$~(see~\eqref{eqn:stop-relays}), and stores  $t_0 +
\tau^{\ast}$ as a header in the packet. \remove{
\begin{enumerate}
\item the time at which copying to relays is stopped - this is set to $t_0 + \tau^{\ast}$.
\item the time at which copying to destinations is stopped - this is set to $t_0 + \tau$.
\end{enumerate}}

The packet is immediately copied to $N_0$ relays, perhaps by means of a
broadcast from an infrastructure "base station". When an infected
node meets a susceptible relay, it compares $t_0 + \tau^{\ast}$ with
the current time. The susceptible relay is not copied to if the
current time exceeds $t_0 + \tau^{\ast}$. However, all the infected
nodes continue to carry the packet, and to copy to susceptible
destinations as and when they meet.

\begin{remarks}
Consider a scenario, where the interest is in copying packet to only
a  fraction $\alpha$ of the destinations. Observe that for
every $\epsilon > 0$,
\begin{equation*}
\lim_{K \rightarrow \infty} \mathbb{P}\left(\left|\frac{m(\tau)}{M}
- \alpha\right| > \epsilon \right) = 0.
\end{equation*}
Thus, in large networks, copying to destinations can also be stopped at
time $\tau$~(see~\eqref{eqn:stop-dstns}) while ensuring that
with large probability the fraction of infected destinations is
close to $\alpha$. Consequently, all the relays can delete the
packet and free their memory at $\tau$.
This helps when packets are large and relay~(cache) memory is limited.
\end{remarks}

\section{Optimal Two-Hop Forwarding}
\label{sec:two-hop}
Instead of epidemic relaying one can consider two-hop
relaying~\cite{comnet-wireless.grossglauser-tse02mobility-adhoc-networks}.
Here, the $N_0$ source nodes can copy the packet
to any of the $N-N_0$ relays or $M$ destinations.
The infected destinations can also copy the packet to
any of the susceptible relays or destinations.
However, the relays are allowed to transmit the packet only
to the destinations. Here also a similar optimization problem as in
Section~\ref{sec:forward-problem} arises.

Now, the decision epochs $t_k, k = 1,2,\dots$ are the meeting epochs of the infected nodes~(sources, relays or destinations)
with the susceptible destinations and the meeting epochs of the sources or infected destinations with the susceptible relays.
We can formulate an MDP with state
\begin{equation*}
s_k := (m_k, n_k, e_k).
\end{equation*}
at instant $t_k$ where $m_k, n_k$ and  $e_k$ are as defined in
Section~\ref{sec:mdp-formulation}. The state space is $[M_{\alpha}-1] \times [N_0:N] \times
 \mathcal{E}$. The control space is
$\mathcal{U} \in \{0,1\}$, where $1$ is for {\it copy} and $0$ is for {\it do not copy}.
We also get a transition structure identical  to that in Section~\ref{sec:mdp-formulation}.

For a state action pair $(s_k,u_k)$
the expected single stage cost is given by
\begin{align*}
g(s_k,u_k) &= \gamma u_k + \mathbb{E}\left\{\delta_{k+1}1_{\{m_{k+1} < M_{\alpha}\}}\right\}\\
           &= \begin{cases}
             \gamma u_k \mbox{ if }  s_k \mbox{ is such that } m_k \geq M_{\alpha} \\
                 \gamma \mbox{ if } s_k  = (M_{\alpha}-1,n,d) \mbox{ and } u_k = 1\\
                 \gamma u_k + C_d(s_k,u_k) \mbox{ otherwise},\end{cases}
\end{align*}
where \\
$C_d(s_k,u_k) = $
\begin{equation*}
\ \ \ \ \ \frac{1}{\begin{split}\big(&(m_k + n_k + u_k)(M-m_k-u_k 1_{\{s_k = d\}})\lambda\\
                                     & + (m_k + u_k1_{\{s_k = d\}} + N_0)(N -n_k-u_k 1_{\{s_k = r\}})\lambda\big)\end{split}}
\end{equation*}
 is the mean time until the next decision epoch.
As before, the quantity $\gamma u_k$ accounts for the
transmission energy.

Let $u^{\ast}: [M_{\alpha}-1] \times [N_0:N] \times \mathcal{E}
\rightarrow \mathcal{U}$ be a stationary optimal policy. As in
Section~\ref{sec:opt-policy}, the optimal policy satisfies
$u^{\ast}(m,n,d) = 1$ for all $(m,n) \in [M-1] \times [N_0:N]$, and
$u^{\ast}(m,n,r) = 0$ for all $(m,n) \in [M_{\alpha}:M-1] \times
[N_0:N]$. Thus, we focus on a reduced state space $[M_{\alpha}-1]
\times [N_0:N] \times \{r\}$. As before, we look for the one step
look ahead policy which turns out to be the same as that for
epidemic relaying. Finally, Theorem~\ref{heu-optimality} holds for
two-hop relaying as well~(see the proof in
Appendix~\ref{proof-heu-optimality}).

Next, we turn to the asymptotically optimal control for two-hop relaying.
The following are the conditional expected
drift rates. For $(m(t),n(t)) \in [M_{\alpha}-1] \times [N_0:N]$,
\begin{align*}
\frac{{\rm d}\mathbb{E}(m(t)|(m(t),n(t)))}{{\rm d}t} = &~\lambda (m(t) + n(t))(M-m(t)), \\
\frac{{\rm d}\mathbb{E}(n(t)|(m(t),n(t)))}{{\rm d}t} = &~\lambda (m(t) + N_0)(N-n(t)) \\
& \ \ \ \ \ \ \ \ \ \ \ \ \ \ \ \ \ \ 1_{\{\Phi(m(t),n(t)) > 0\}}.
\end{align*}
We employ the same scaling and notations as in~\eqref{eqn:scalings}. The drift rates
in terms of $(x^K(t),y^K(t)) \in [0, X_{\alpha}] \times [Y_0,Y]$ are
\begin{align*}
\frac{{\rm d}\mathbb{E}(x^K(t)|(x^K(t),y^K(t)))}{{\rm d}t} &= f^K_1(x^K(t),y^K(t))  \\
                                               := &~\Lambda (x^K(t) + y^K(t))(X-x^K(t)), \\
\frac{{\rm d}\mathbb{E}(y^K(t)|(x^K(t),y^K(t)))}{{\rm d}t} &= f^K_2(x^K(t),y^K(t)) \\
                                               := \Lambda (x^K(t) + Y_0)&(Y-y^K(t))1_{\{\phi^K(x^K(t),y^K(t)) > 0\}},
\end{align*}
Now, $x(t),y(t)$ are defined as functions satisfying $x(0)= 0, y(0) = Y_0$ and for
$t \geq 0$,
\begin{align*}
\frac{{\rm d}x(t)}{{\rm d}t} = &~f_1(x(t),y(t)) := \Lambda (x(t) + y(t))(X-x(t)), \\
\frac{{\rm d}y(t)}{{\rm d}t} = &~f_2(x(t),y(t)) := \Lambda (x(t) + Y_0)(Y-y(t)) \\
                            & \ \ \ \ \ \ \ \ \ \ \ \ \ \ \ \ \ \ \ \ \ \ \ \ \ \ \ \ \ \ \ \ \ \ \ \ \ \ \ 1_{\{\phi(x(t),y(t)) > 0\}}
\end{align*}
The analysis in Section~\ref{asym-opt-forward} applies to two-hop relaying
as well. In particular, Theorems~\ref{assym-optimality} and~\ref{theorem:asym-optimal} hold.
However, for the identical system parameters~($M,N,\alpha,\lambda$ and $\gamma$)
and initial state~($N_0$), the value of the time-threshold $\tau^{\ast}$ will be larger on account of the
slower rates of infection of relays and destinations.

\begin{figure}[t]
\centering
\includegraphics[width=2.6in]{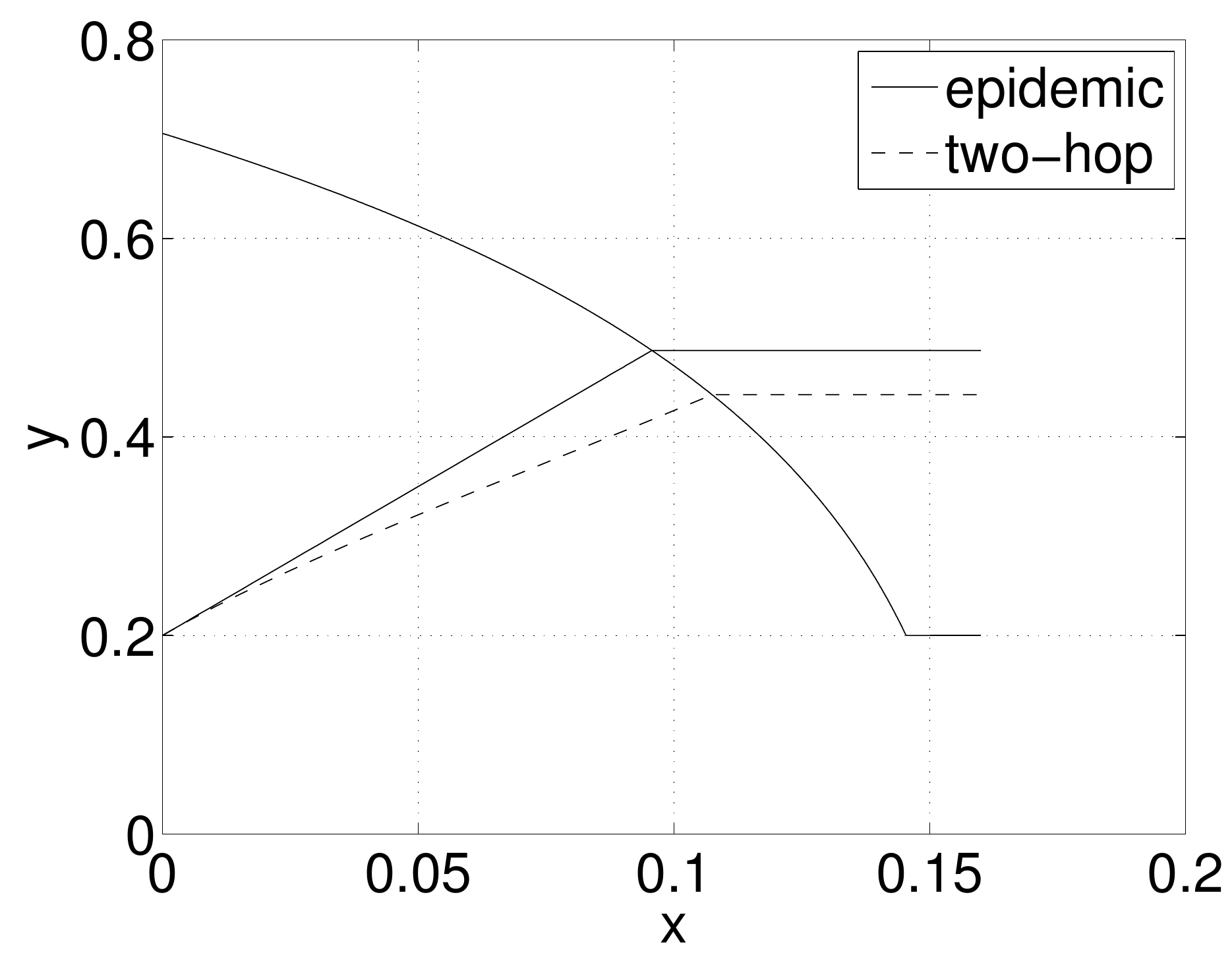}
\caption{An illustration of the  epidemic and two hop trajectories.
 The plots also show the graph of  `$\phi(x,y) = 0$'.}
\label{fig:epi-twohop-trajectory}
\end{figure}

\begin{figure}[t]
\centering
\subfigure{\includegraphics[width=2.6in]{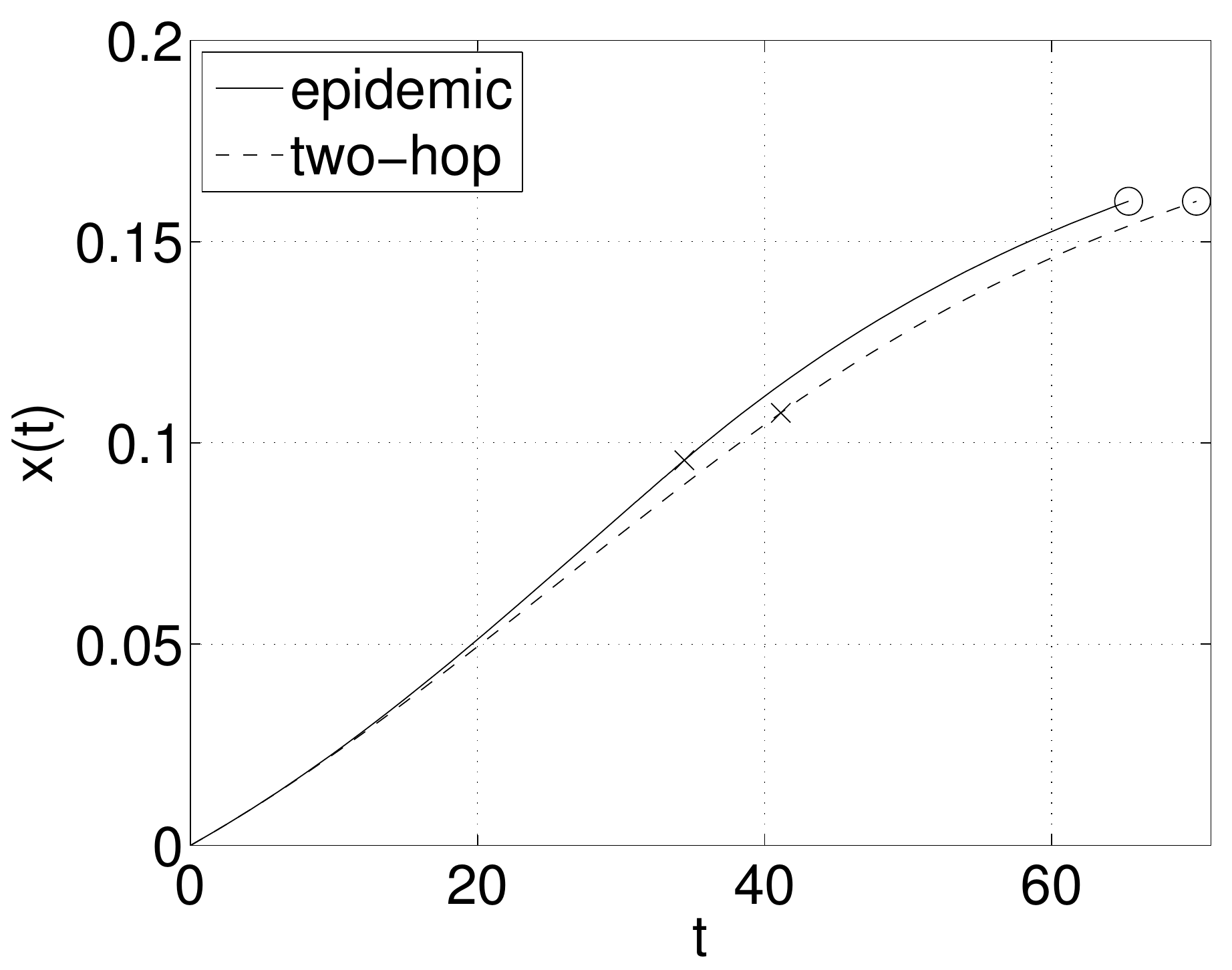}}
\subfigure{\includegraphics[width=2.6in]{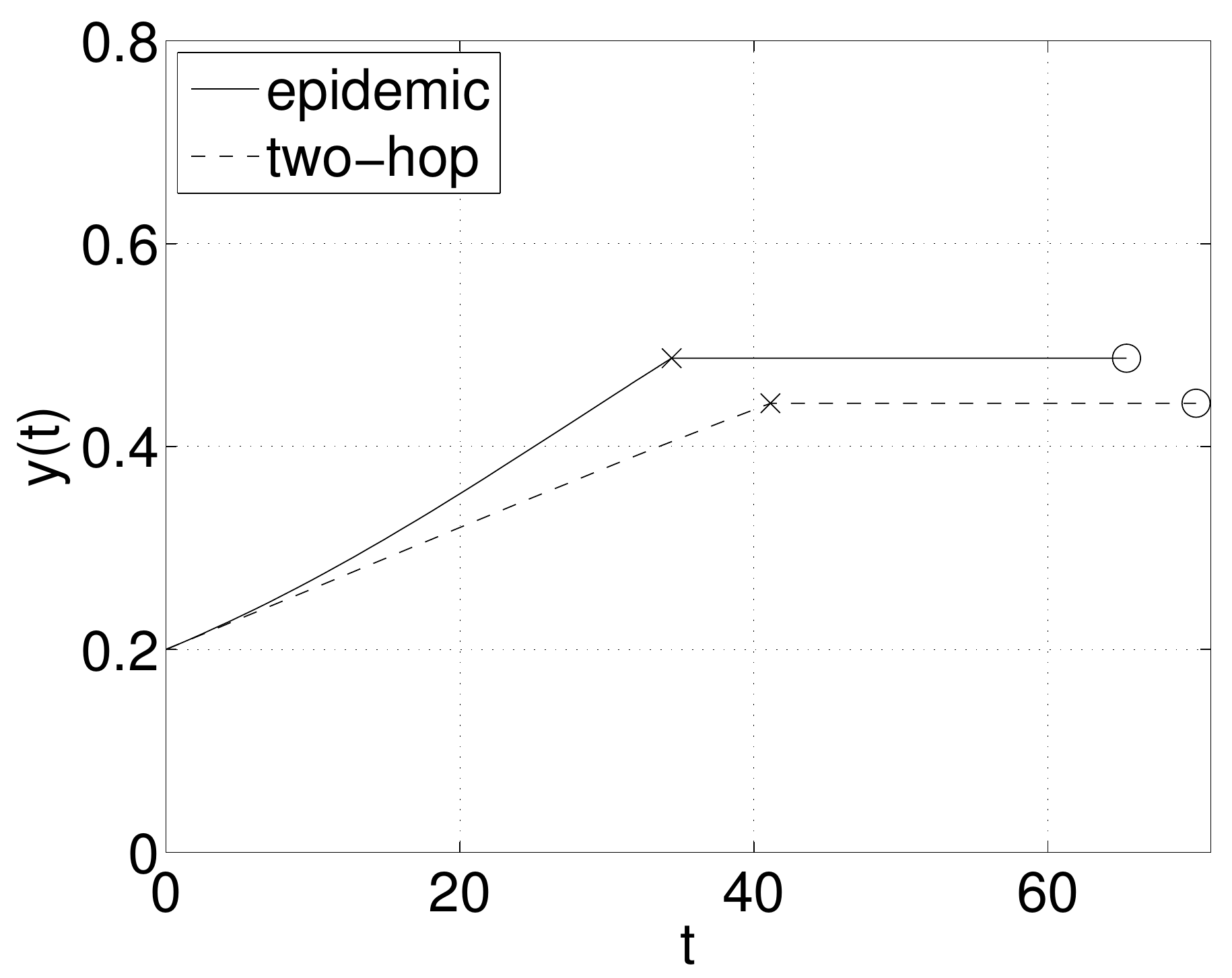}}
\caption{The top and bottom sub-plots respectively show the fractions of infected
destinations and relays as a function of time. The marker 'X' indicates the states at
which copying to relays is stopped, and 'O' indicates the states at
which $\alpha$ fraction of destinations have been copied.}
\label{fig:epi-twohop-evolution}
\end{figure}

We illustrate the comparison between epidemic and two-hop relaying using an example.
Let $X = 0.2,Y = 0.8,\alpha = 0.8,Y_0 = 0.2,\Lambda = 0.05$
and $\Gamma = 50$. In Figure~\ref{fig:epi-twohop-trajectory}, we plot the
graph of  ``$\phi(x,y) = 0$'', and also the '$y$ versus $x$' trajectories corresponding to
 epidemic and two-hop relayings. In Figure~\ref{fig:epi-twohop-evolution},
we plot the trajectories of $(x(t),y(t))$ corresponding to
 epidemic and two-hop relayings. As anticipated, the value of the time-threshold $\tau^{\ast}$ is larger
for two-hop relaying than epidemic relaying. Moreover, the number of transmissions is less while the deliverly delay is more 
under the controlled two-hop relaying. 

\section{Numerical Results}
\label{num-results}
\begin{figure}[t]
\centering
\subfigure{\includegraphics[width=2.8in]{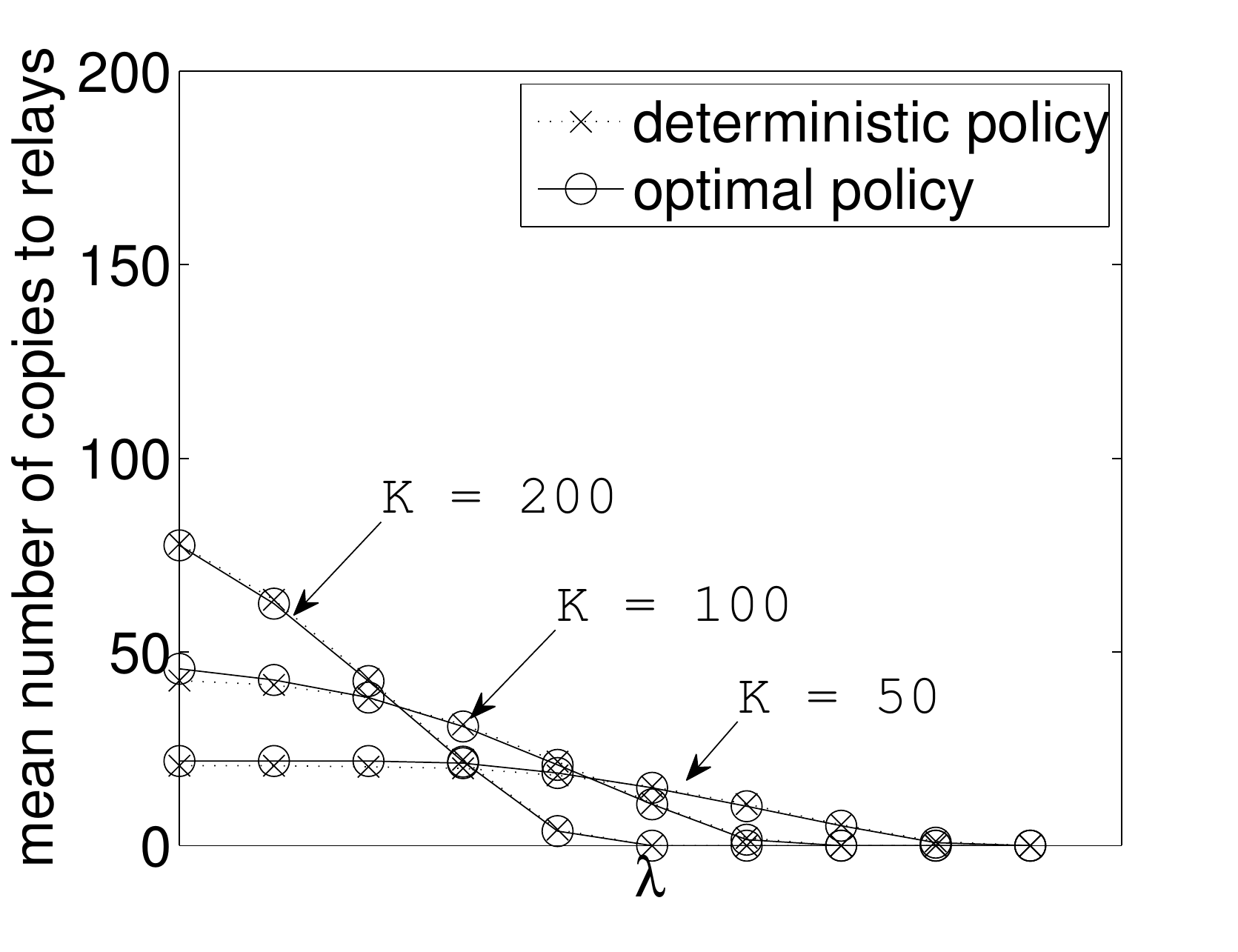}}
\subfigure{\includegraphics[width=2.8in]{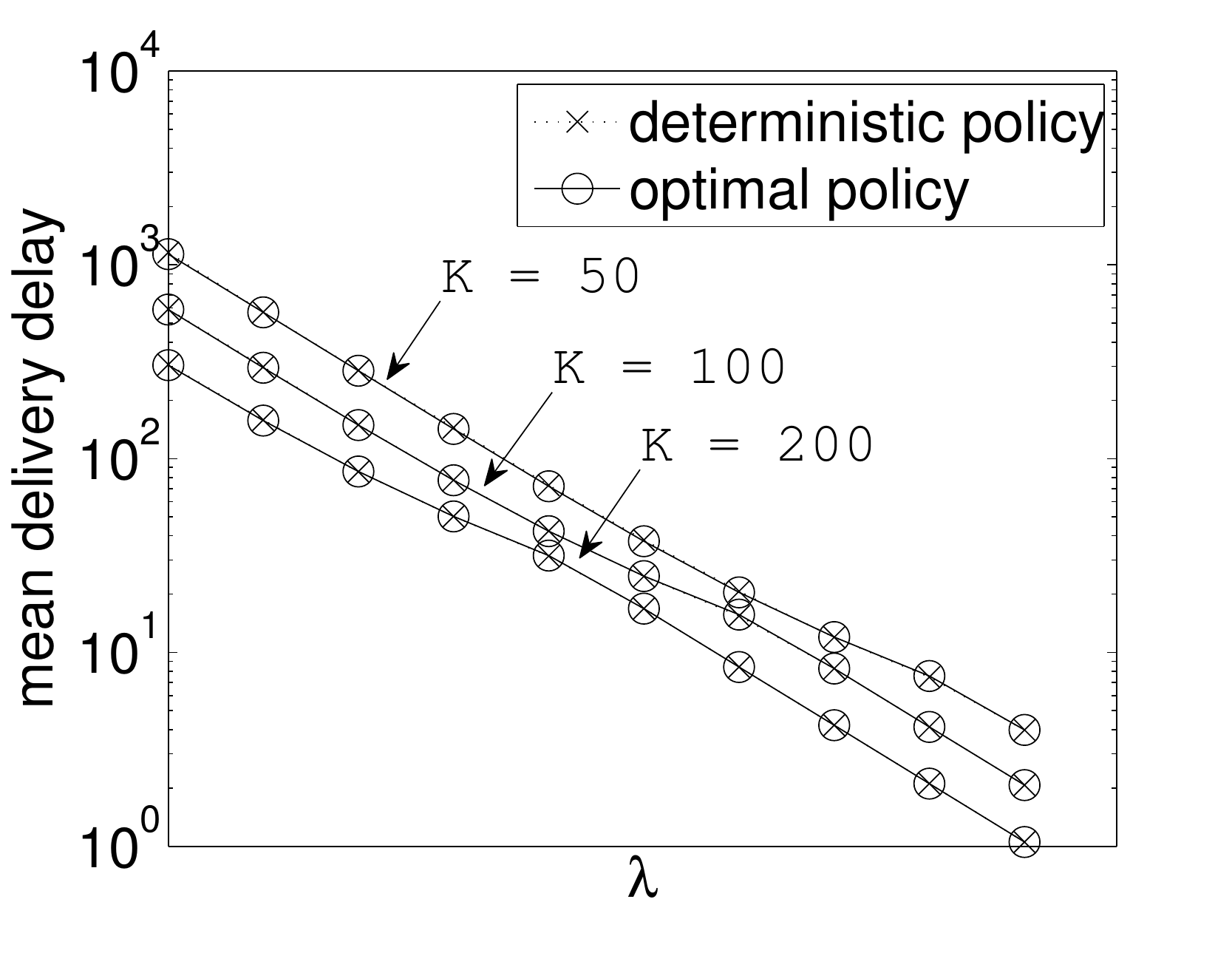}}
\caption{The top and bottom sub-plots, respectively, show the total
number of copies to relays and the delivery delays corresponding to
both the optimal and the deterministic policies.}
\label{fig:numerical}
\end{figure}

We now show some numerical results to demonstrate the good performance of the deterministic control
in epidemic forwarding in a DTN with multiple destinations.
Let $X = 0.2,Y = 0.8,\alpha = 0.8,Y_0 = 0.2$ and $\gamma = 0.5$.
We vary $\lambda$ from $0.00005$ to $0.05$ and use $K = 50,100$ and $200$.
In Figure~\ref{fig:numerical}, we plot the total number of copies to relays and the delivery delays corresponding
to both the optimal and the  asymptotically optimal deterministic policies.
Evidently, the deterministic policy
performs close to the optimal policy on both the fronts.
We observe that, for a fixed $K$, both the mean delivery delay and the
 mean number of copies to relays decrease as $\lambda$ increases.
We also observe that, for a fixed $\lambda$, the mean delivery delay decreases
as the network size grows.
Finally, for smaller values of $\lambda$, the mean number of copies to relays
increases  with the network size, and for larger values of $\lambda$, the opposite happens.

\section{Conclusion}
We studied the epidemic forwarding in DTNs, formulated the problem as a controlled continuous time Markov chain, and
obtained the optimal policy~(Theorem~\ref{heu-optimality}). We then developed an ordinary differential equation
approximation for the optimally controlled Markov chain, under a natural scaling, as the population of nodes
increases to $\infty$~(Theorem~\ref{assym-optimality}). This o.d.e. approximation yielded a forwarding policy that does not require
global state information (and, hence, is implementable), and is asymptotically optimal~(Theorem~\ref{theorem:asym-optimal}).

The optimal forwarding problem can also be addressed following the
result of Gast et al.~\cite{stochctrl.gast-etal10mean-field-MDPs}.
They study a general discrete time Markov decision process~(MDP)~\cite{stochctrl.bertsekas07dpoc-vol2}.
However, they do not solve the finite problem citing the difficulties
associated with obtaining the asymptotics of the optimally controlled
process~(see~\cite[Section~3.3]{stochctrl.gast-etal10mean-field-MDPs}).
Instead, they consider the fluid limit
of the MDP, and analyze optimal control over the deterministic limiting problem.
They then show that the optimal reward of the MDP converges to the optimal reward of its
mean field approximation, given by the solution to a Hamilton-Jacobi-Bellman~(HJB)
equation~\cite[Section~3.2]{stochctrl.bertsekas05dpoc-vol1}.
On the other hand, our approach is more direct.
We have a continuous time controlled Markov chain at our disposal
We explicitly characterize the optimal policy for the finite~(complete information) problem,
and prove convergence of the optimally controlled Markov chain to a fluid limit.
An asymptotically optimal deterministic control is then suggested by the limiting deterministic dynamics,
and does not require solving HJB equations. Our notion of asymptotic optimality is also stronger in the
sense that we apply both the optimal policy and the deterministic policy to the finite problem, and show that
the corresponding costs converge.

There are several directions in which this work can be extended.
In the same DTN framework, there could be a deadline on the delivery time of the packet~(or message);
the goal of the optimal control could be to maximize the fraction of destinations that receive
the packet before the deadline subject to an energy constraint. Our work in this paper
assumes that network parameters such as $M,N,\lambda$ etc., are known;
it will be important to address the adaptive control problem when these parameters are unknown.

\appendices
\section{Proof of Theorem~\ref{heu-optimality}}
\label{proof-heu-optimality}

We first prove that for the optimal policy it is sufficient to consider two
actions $1$~(i.e., copy now) and {\it stop}~(i.e., do not copy now and never copy again).
More precisely, under the optimal policy, if a susceptible relay
that is met is not copied, then no susceptible relay is copied in the future as well.
Let us fix a $N_0 \leq n \leq N-1$. Let  $m^{\ast}_n$ be the maximum $j$
such that $u^{\ast}(j,n,r) = 1$.\footnote{Note that, for a given $n$,  $m^{\ast}_n$ could be 0,
in that case we do not copy to any more relays.}
We show that $u^{\ast}(j,n,r) = 1$ for all $0 \leq j < m^{\ast}_n$; see Figure~\ref{fig:heu-policy}
for an illustration of this fact.
The proof is via induction.
\remove{
Let us assume that $u^{\ast}(j,n,r) = 1$ for all $m + 1 \leq j
\leq m^{\ast}_n$. The following results completes the induction step.
}

\begin{proposition}
If $u^{\ast}(j,n,r) = 1$ for all $m+1 \leq j \leq m^{\ast}_n$, then $u^{\ast}(m,n,r) = 1$.
\end{proposition}
\begin{IEEEproof}
Define
\begin{align*}
\psi(m,n) := &~J_{0s}(m,n,r) - J(m,n,r),\\
\theta_0(m,n) := &~J_{0s}(m,n,r) - A((m,n,r),0), \\
\mbox {and } \theta_1(m,n) := &~J_{1s}(m,n,r) - A((m,n,r),1).
\end{align*}
Both the action sequences that give rise to the two cost terms in
the definition of $\theta_0(m,n)$, do not copy to the
susceptible relay that was just met. Let $j$ be the number of infected
destinations at the next decision epoch when a susceptible relay is
met; $j$ can be $m, m+1, ..., M$. All interim decision epochs must be
meetings with susceptible destinations, and both policies copy at
these meetings. Hence, both policies incur the same cost until this
epoch, and differ by $\psi(j,n)$ in the costs to go~(from this epoch onwards).
Averaging the difference over $j$, and noting that $\psi(j,n) = 0$ for $j >
M_{\alpha} -1$, we get\footnote{We use the
standard convention that a product over an empty index set is $1$,
which happens when $j = m$.}
\begin{equation}
\label{eqn:theta_0}
 \theta_0(m,n) = \sum_{j =m}^{M_{\alpha}-1}\left(\prod_{l =m}^{j-1}p_{l,n}(d)\right)p_{j,n}(r)\psi(j,n).
\end{equation}
Since $A((m,n,r),0) \geq J(m,n,r)$, it follows that $\psi(m,n) \geq \theta_0(m,n)$, and so
\vspace{1pt}
\begin{align*}
 \psi(m,n) \geq & \sum_{j = m}^{M_{\alpha}-1}\left(\prod_{l =
m}^{j-1}p_{l,n}(d)\right)p_{j,n}(r)\psi(j,n)  \\
= & \ p_{m,n}(r)\psi(m,n) \\
& +  p_{m,n}(d) \sum_{j = m+1}^{M_{\alpha}-1}\left(\prod_{l =
m+1}^{j-1}p_{l,n}(d)\right)p_{j,n}(r)\psi(j,n)  \\
\end{align*}
which implies upon rearrangement
\begin{equation}
\psi(m,n) \geq \sum_{j = m+1}^{M_{\alpha}-1}\left(\prod_{l =
m+1}^{j-1}p_{l,n}(d)\right)p_{j,n}(r)\psi(j,n)   \label{eqn:psi-nm}
\end{equation}

Next, we establish the following lemma.
\begin{lemma}
\label{lma:lemma2}
$\theta_1(m,n) \geq \theta_1(m+1,n).$
\end{lemma}
\begin{IEEEproof}
Note that both the action sequences that lead to the two cost terms in
the definition of $\theta_1(m,n)$ copy at state $(m,n,r)$.
Subsequently, both incur equal costs until a decision epoch when
an infected node meets a susceptible relay. Also, at any such state
$(j,n+1,r),~ j \geq m$, the costs to go differ by $\psi(j,n+1)$.
Hence,
\begin{align*}
\theta_1(m,n) &= \sum_{j = m}^{M_{\alpha}-1}\left(\prod_{l =
m}^{j-1}p_{l,n+1}(d)\right)p_{j,n+1}(r)\psi(j,n+1)\\
 &= p_{m,n+1}(r)\psi(m,n+1) + p_{m,n+1}(d)\theta_1(m+1,n)
\end{align*}
where
\begin{equation*}
\theta_1(m+1,n)= \sum_{j = m+1}^{M_{\alpha}-1}\left(\prod_{l = m+1}^{j-1}p_{l,n+1}(d)\right)p_{j,n+1}(r)\psi(j,n+1).
\end{equation*}
Thus it suffices to show that
\begin{equation*}
\psi(m,n+1) \geq \theta_1(m+1,n).
\end{equation*}
which is same as~\eqref{eqn:psi-nm} with $n$ replaced by $n+1$.
\end{IEEEproof}

Next, observe that for all $m \leq j \leq m^{\ast}_n$,
\begin{align}
\psi(j,n) &~= J_{0s}(j,n,r) - \min\{A((j,n,r),0),A((j,n,r),1)\} \nonumber \\
  &~= \max\{\theta_0(j,n),\Phi(j,n) + \theta_1(j,n)\}. \label{eqn:psi-jn}
\end{align}
Moreover, from the induction hypothesis, the optimal policy copies at states
$(j,n,r)$ for all $m+1 \leq j \leq m^{\ast}_n$. Hence, for $m+1 \leq j \leq m^{\ast}_n$,
\begin{equation*}
\psi(j,n) = \Phi(j,n) + \theta_1(j,n).
\end{equation*}
Finally, $\psi(j,n) = 0$ for all $m^{\ast}_n < j \leq M_{\alpha}-1$ as the optimal policy does not copy
in these states.
Hence, from~\eqref{eqn:theta_0},
\begin{align}
\lefteqn{\theta_0(m,n)} \nonumber \\
&~=  p_{m,n}(r)\max\{\theta_0(m,n),\Phi(m,n) + \theta_1(m,n)\}+ p_{m,n}(d) \nonumber \\
&\ \ \ \ \ \ \ \times \sum_{j = m+1}^{m^{\ast}_n}\left(\prod_{l = m+1}^{j-1}p_{l,n}(d)\right)p_{j,n}(r)\big(\Phi(j,n) +\theta_1(j,n)\big) \nonumber \\
&~< p_{m,n}(r)\max\left\{\theta_0(m,n),\Phi(m,n) + \theta_1(m,n)\right\}+ p_{m,n}(d) \nonumber \\
&\ \ \ \  \times \big(\Phi(m,n) + \theta_1(m,n)\big) \sum_{j = m+1}^{m^{\ast}_n}\left(\prod_{l = m+1}^{j-1}p_{l,n}(d)\right)p_{j,n}(r)\nonumber \\
&~\leq p_{m,n}(r)\max\left\{\theta_0(m,n),\Phi(m,n) + \theta_1(m,n)\right\}\nonumber \\
&\ \ \ \ + p_{m,n}(d) \big(\Phi(m,n) + \theta_1(m,n)\big) \nonumber \\
&~=\max\left\{p_{m,n}(r)\theta_0(m,n) + p_{m,n}(d) \big(\Phi(m,n) + \theta_1(m,n)\big),\right. \nonumber \\
&\ \ \ \ \ \ \ \ \ \  \ \ \left. \Phi(m,n) + \theta_1(m,n)\right\}, \label{eqn:bound-2}
\end{align}
where the first~(strict) inequality holds because $\Phi(m,n)$ is strictly decreasing~(see~\eqref{eqn:Phi})
and $\theta_1(m,n)$ is decreasing~(see Lemma~\ref{lma:lemma2}) in $m$ for fixed
$n$. The second inequality follows because the summation term is a probability which is less than $1$.
Now suppose that $\theta_0(m,n) \geq  \Phi(m,n) + \theta_1(m,n)$.
Then
\begin{align*}
\lefteqn{\max\left\{p_{m,n}(r)\theta_0(m,n) + p_{m,n}(d) \big(\Phi(m,n) + \theta_1(m,n)\big),\right.}\\
& \ \ \ \ \ \  \ \ \left. \Phi(m,n) + \theta_1(m,n)\right\}\\
&~= p_{m,n}(r)\theta_0(m,n) + p_{m,n}(d) \big(\Phi(m,n) + \theta_1(m,n)\big)\\
&~\leq  \theta_0(m,n)
\end{align*}
which contradicts~\eqref{eqn:bound-2}. Thus, we conclude that
\begin{equation*}
\theta_0(m,n) <  \Phi(m,n) + \theta_1(m,n).
\end{equation*}
This further implies that $\psi(m,n) = \Phi(m,n) + \theta_1(m,n)$ (see~\eqref{eqn:psi-jn}),
and so that $u^{\ast}(m,n,r) = 1$.
\end{IEEEproof}

We now return to the proof of Theorem~\ref{heu-optimality}.
We show that the one-step look ahead policy is optimal for the resulting
stopping problem. To see this, observe that $\Phi(m,n)$ is decreasing in $m$ for
a given $n$ and also decreasing in $n$ for a given $m$. Thus,
if $(m,n,r) \in \mathcal{S_S}$, i.e, $\Phi(m,n) \leq 0$~(see~\eqref{eqn:stopping-set}), and the
susceptible relay that is met is copied, the next state $(m,n+1,r)$ also belongs to the
stopping set $\mathcal{S_S}$. In  other words,  $\mathcal{S_S}$ is also an
absorbing set~\cite[Section~3.4]{stochctrl.bertsekas07dpoc-vol2}).
Consequently, the one-step look ahead policy is an optimal policy.

\section{Proof of Theorem~\ref{assym-optimality}}
\label{proof-assym-optimality}
We start with a preliminary result and a few definitions.
\begin{proposition}
\label{prop-uni-conv} Let $\alpha < 1$ and $Y_0 > 0$. Let $\phi^K$
and $\phi$ be as given in~\eqref{eqn:phi-K} and~\eqref{eqn:phi},
respectively. Then, the functions $\phi^K(\cdot)$ converge to
$\phi(\cdot)$ uniformly, i.e., for every $\nu > 0$, there exists a
$K_{\nu}$ such that
\begin{equation*}
\sup_{(x,y) \in \Delta^K} |\phi^K(x,y) - \phi(x,y)| < \nu
\end{equation*}
for all $K \geq K_{\nu}$.
\end{proposition}
\begin{IEEEproof}
For a $y \in [Y_0, Y]$, define $f_y:[0, X_{\alpha}]
\rightarrow \mathbb{R}_+$ as follows.
\[
f_y(z) = \frac{1}{(y+z)^2(X-z)}.
\]
Clearly, the family $\{f_y\}$ is positive and uniformly upper
bounded. Indeed,
\[f_y(z) \leq f_{\max} := \frac{1}{Y_0^2(X-X_{\alpha})}.\]
Further,
\[
\frac{{\rm d}f_y(z)}{{\rm d}z} =
\frac{1}{(y+z)^2(X-z)}\left(\frac{1}{X-z}-\frac{2}{y+z}\right),
\]
from which it can be seen that
\[\left|\frac{{\rm d}f_y(z)}{{\rm d}z}\right| \leq f'_{\max}\]
where $f'_{\max}$ is a suitably defined constant. So the family
$\{f_y\}$ is uniformly Lipschitz. Now, for $(z,y) \in [0,
X_{\alpha}] \times [Y_0, Y]$,
\begin{align}
&\frac{1}{K(y+z)(y+z+\frac{1}{K})(X-z)}  - \int_z^{z + \frac{1}{K}}f_y(v){\rm d}v \nonumber \\
&~\leq \frac{f_y(z)}{K} - \int_z^{z +\frac{1}{K}}f_y(v){\rm d}v \nonumber \\
&~\leq \int_z^{z +\frac{1}{K}}(f_y(z) - f_y(v)){\rm d}v \nonumber \\
&~\leq \frac{f'_{\max}}{K^2} \label{inq-left}
\end{align}
where the first and the last inequalities follow from the
definitions of $f_y(z)$ and $f'_{\max}$ respectively. On the other
hand,
\begin{align*}
\frac{1}{(y+z)(y+z+\frac{1}{K})(X-z)} &~= f_y(z)\frac{y+z}{y+z+\frac{1}{K}} \\
&~\geq  f_y(z) \frac{Y_0}{Y_0 + \frac{1}{K}}.
\end{align*}
Hence
\begin{align}
&\int_z^{z + \frac{1}{K}}f_y(v){\rm d}v - \frac{1}{K(y+z)(y+z+\frac{1}{K})(X-z)} \nonumber \\
&~\leq \int_z^{z +\frac{1}{K}}f_y(v){\rm d}v - \frac{f_y(z)}{K} \frac{K Y_0}{1  + K Y_0} \nonumber \\
&~\leq \int_z^{z +\frac{1}{K}}(f_y(v) - f_y(z)){\rm d}v + \frac{f_y(z)}{K(1  + K Y_0)} \nonumber \\
&~ \leq \frac{f'_{\max}}{K^2} + \frac{f_{\max}}{K(1 + KY_0)}
\label{inq-right}.
\end{align}
Combining~\eqref{inq-left} and~\eqref{inq-right},
\begin{align*}
&\left|\frac{1}{K(y+z)(y+z+\frac{1}{K})(X-z)} - \int_z^{z + \frac{1}{K}}f_y(v){\rm d}v \right| \\
&~\leq \frac{f'_{\max}}{K^2} + \frac{f_{\max}}{K(1 + KY_0)}.
\end{align*}
Now fix a $(x,y) \in \Delta^K$. Setting $z = j/K$, and summing over
$j \in [Kx:\lceil KX_{\alpha} \rceil -1]$, we get
\begin{align*}
\lefteqn{|\phi^K(x,y) - \phi(x,y)|} \\
&~\leq \sum_{j = Kx}^{\lceil KX_{\alpha} \rceil -1}
\frac{1}{\Lambda}\left| \frac{1}{K(y + \frac{j}{K})(y +
\frac{j+1}{K})(X
-\frac{j}{K})}\right. \\
&~ \ \ \ \ \ \ \ \ \ \ \ \ \ \ \ \ \ \ \left. - \int_{\frac{j}{K}}^{\frac{j+1}{K}}f_y(v){\rm d}v \right| + \frac{1}{\Lambda}\left| \int_{\frac{\lceil KX_{\alpha} \rceil}{K}}^{X_{\alpha}}f_y(v)dv \right| \\
&~\leq \frac{K(X_{\alpha}-x)}{\Lambda}\left(\frac{f'_{\max}}{K^2} + \frac{f_{\max}}{K(1 + KY_0)}\right) + \frac{f_{\max}}{K \Lambda} \\
&~\leq \frac{X_{\alpha}f'_{\max}}{K\Lambda} +
\frac{X_{\alpha}f_{\max}}{(1 + K Y_0)\Lambda} +
\frac{f_{\max}}{K\Lambda}.
\end{align*}
The obtained upper bound on the right-hand side is independent of
$(x,y) \in \Delta^K$, and vanishes as $K
\rightarrow \infty$. Thus, for every $\nu > 0$, there exists a
$K_{\nu}$ such that
\begin{equation*}
\sup_{(x,y) \in \Delta^K}|\phi^K(x,y) - \phi(x,y)| < \nu
\end{equation*}
for all $K \geq K_{\nu}$.
\end{IEEEproof}

In the following, to facilitate a parsimonious description, we use
the notation $z^K(t) = (x^K(t),y^K(t))$, $z(t) = (x(t),y(t))$ and
$\mathcal{Z} = [0, X_{\alpha}] \times [Y_0, Y]$. Let us define, for
a $\nu \in \mathbb{R}$,
\begin{align*}
\mathcal{S}_{\nu} &~= \{z \in  \mathcal{Z}: \phi(z) > \nu \}, \\
\tau_{\nu} &~= \inf\{t \geq 0:z(t) \notin \mathcal{S}_{\nu}\},
\end{align*}
and a stopping time
\[
\tau^K_{\nu} = \inf\{t \geq 0:z^K(t) \notin \mathcal{S}_{\nu}\},
\]
the time when $z^K(t)$ exits the limiting set $\mathcal{S}_\nu$.
Observe that
\begin{equation}
\frac{\partial \phi}{\partial x} = -\frac{1}{\Lambda(x+y)^2 (X-x)}
\leq -\frac{1}{\Lambda(X_{\alpha}+Y)^2 X} \label{eqn:rate-phi}
\end{equation}
 and $f^K_1(x,y)$ defined in~\eqref{eqn:f-K_1} is positive and is also bounded away from zero.
 These imply that $\tau^K_{\nu} < \infty$ with probability $1$.
Similarly, $\tau_{\nu} < \infty$. The following assertion is a
corollary of Proposition~\ref{prop-uni-conv}.
\begin{corollary}
\label{cor:apprx-bdry} Let $K_{\nu}$ be as in
Proposition~\ref{prop-uni-conv}. For $K \geq K_{\nu}$,
\begin{align*}
\phi^K(z) > 0 &~\mbox{for all } z \in \mathcal{S}_{\nu},\\
\mbox{and } \phi^K(z) \leq 0 &~\mbox{for all } z \notin
\mathcal{S}_{-\nu}.
\end{align*}
\end{corollary}

We define the uncontrolled dynamics~(i.e., the one in which the
susceptible relays are always copied) as a Markov process
$\bar{z}^K(t) = (\bar{x}^K(t),\bar{y}^K(t))$, $t \geq 0$ for which
$\bar{z}^K(0) = z^K(0)$. Let $\bar{z}(t) = (\bar{x}(t),\bar{y}(t))$,
$t \geq 0$ be the corresponding limiting deterministic dynamics.
Formally, $\bar{z}(0) = z(0)$, and for $t \geq 0$,
\begin{align*}
\frac{{\rm d}\bar{x}(t)}{{\rm d}t} = \Lambda (\bar{x}(t) + \bar{y}(t))(X-\bar{x}(t)), \\
\frac{{\rm d}\bar{y}(t)}{{\rm d}t} = \Lambda (\bar{x}(t) +
\bar{y}(t))(Y-\bar{y}(t)).
\end{align*}
The quantities on the right-hand side of the above equations are at
most $\Lambda$, and so
\[
\left \Vert \frac{{\rm d}\bar{z}}{{\rm d}t}\right \Vert \leq
\sqrt{2}\Lambda.
\]
Also observe that the processes $\bar{z}^K(t)$ and $\bar{z}(t)$
satisfy the hypotheses of Darling~\cite{stochproc.darling02fluid-limits}~(see Section~\ref{asym-det-dynamics}),
and thus convergence of $\bar{z}^K(t)$ to $\bar{z}(t)$ follows.

We also define a Markov process $\tilde{z}^K(t) =
(\tilde{x}^K(t),\tilde{y}^K(t))$, $t \geq \tau_{\nu}$ for which
$\tilde{z}^K(\tau_{\nu}) = z^K(\tau_{\nu})$ and
\begin{align*}
\frac{{\rm
d}\mathbb{E}(\tilde{x}^K(t)|(\tilde{x}^K(t),\tilde{y}^K(t))}{{\rm
d}t} =&~\Lambda (\tilde{x}^K(t) + \tilde{y}^K(t))(X-\bar{x}^K(t))\\
\frac{{\rm d}\mathbb{E}(y^K(t)|(x^K(t),y^K(t))}{{\rm d}t} = &~0 \\
\end{align*}
In other words, $\tilde{z}^K(t)$ is the process in which relays are
not copied from $\tau_{\nu}$ onwards. Similarly, we define
$\tilde{z}(t) = (\tilde{x}(t),\tilde{y}(t))$, $t \geq \tau_{\nu}$ as
the solution of the corresponding differential equations. In other
words, $\tilde{z}(\tau_{\nu}) = z(\tau_{\nu})$, and for $t \geq
\tau_{\nu}$,
\begin{align*}
\frac{{\rm d}\tilde{x}(t)}{{\rm d}t} = &~f_1(\tilde{x}(t),\tilde{y}(t)) := \Lambda (\tilde{x}(t) + \tilde{y}(t))(X-\tilde{x}(t)), \\
\frac{{\rm d}\tilde{y}(t)}{{\rm d}t} =
&~f_2(\tilde{x}(t),\tilde{y}(t)) := 0
\end{align*}
 We define
\begin{align*}
\tilde{\tau}^K_{-\nu} &~= \inf\{t \geq \tau_{\nu}: \tilde{z}^K(t)
\notin
\mathcal{S}_{-\nu}\},\\
\tilde{\tau}_{-\nu} &~= \inf\{t \geq \tau_{\nu}: \tilde{z}(t) \notin
\mathcal{S}_{-\nu}\}.
\end{align*}
Since
\[\Lambda Y_0(X-X_{\alpha}) \leq \frac{{\rm d}\tilde{x}}{{\rm d}t} \leq \Lambda,\]
the lower bound implies that there is a strictly positive increase
in $\tilde{x}$ after time $\tau_{\nu}$. Since $\Phi(x, y)$ decreases
with increasing $x$ at a rate bounded away from
$0$~(see~\ref{eqn:rate-phi}), $\tilde{z}(t)$ must exit $S_{-\nu}$
within a short additional duration. Thus, we have that
\[\tilde{\tau}_{-\nu} - \tau_{\nu} \leq b \nu\]
for a suitably chosen $b < \infty$.

\begin{figure}[b]
\centering
\includegraphics[width=3.6  in]{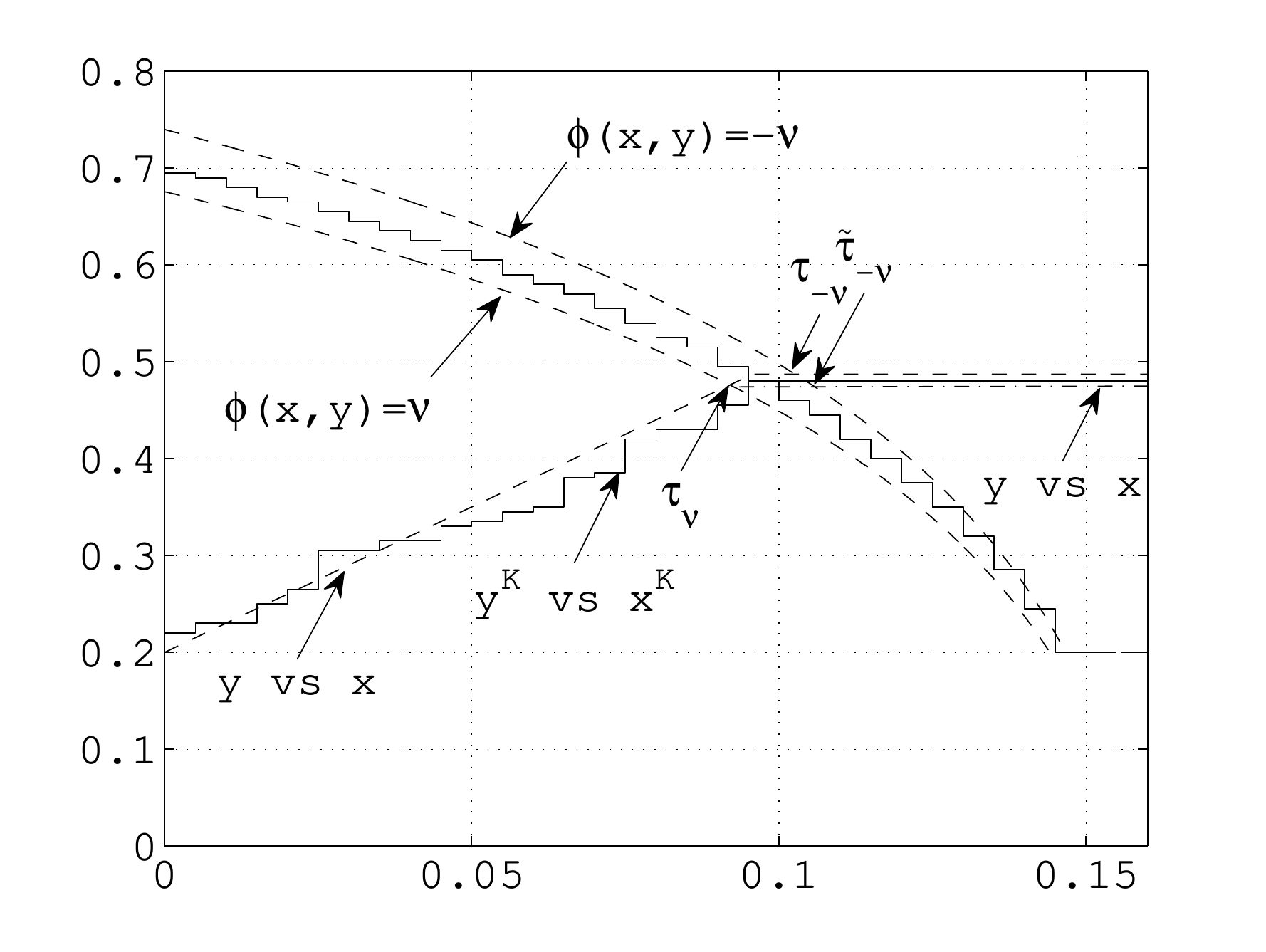}
\caption{An illustration of the trajectories of the controlled CTMC and the corresponding ODE, and the associated variables.}
\label{fig:boundries}
\end{figure}

To aid the reader, we summarize
the variables used in Table~\ref{variables-description}.
We also illustrate sample trajectories of a controlled CTMC and the corresponding
ODE via an example~(Figure~\ref{fig:boundries}). We choose $M = 40, N = 160, \alpha = 0.8, N_0 = 40, \lambda = 0.00025$
and $\gamma = 0.25$. We plot the graphs of '$\phi(x,y) = \nu$' and '$\phi(x,y) = -\nu$'
for $\nu = 0.2$. We also show the trajectories ``$y^K$ vs $x^K$'', ``$y$ vs $x$'',  ``$\tilde{y}$ vs $\tilde{x}$''
and the epochs $\tau_{\nu}$, $\tau_{-\nu}$ and $\tilde{\tau}_{-\nu}$.

\begin{table}[t]
\renewcommand{\arraystretch}{1.3}
\caption{Variables and their description}
\label{variables-description}
\centering
\begin{tabular}{l|l}
\hline
\bfseries variables & \bfseries description \\
\hline
 $z^K(t)$ & controlled dynamics with discontinuity at $\tau^K$ \\
 $z(t)$ & $z^K(t)$'s fluid limit with discontinuity at $\tau^{\ast}$ \\
 $\tau^K_{\nu}$ & instant when $z^K(t)$ exits $\mathcal{S}_{\nu}$ \\
 $\tau_{\nu}$ & instant when $z(t)$ exits $\mathcal{S}_{\nu}$ \\
 $\bar{z}^K(t)$ & uncontrolled dynamics with no discontinuity \\
 $\bar{z}(t)$ & $\bar{z}^K(t)$'s fluid limit with no discontinuity \\
 $\tilde{z}^K(t)$ & identical to $z^K(t)$ until $\tau_{\nu}$ at which
 copying to \\
        & relays is stopped \\
 $\tilde{z}(t)$ & $\tilde{z}^K(t)$'s fluid limit with discontinuity at $\tau_{\nu}$ \\
 $\tilde{\tau}^K_{-\nu}$ & instant when $\tilde{z}^K(t)$ exits $\mathcal{S}_{-\nu}$ \\
 $\tilde{\tau}_{-\nu}$ & instant when $\tilde{z}(t)$ exits $\mathcal{S}_{-\nu}$ \\
\hline
\end{tabular}
\end{table}

We prove the assertion in Theorem~\ref{assym-optimality} in three
steps:
\begin{inparaenum}[(a)]
\item over $[0, \tau_{\nu}]$,
\item over $[ \tau_{\nu}, \tilde{\tau}_{-\nu}]$ and
\item over $[\tilde{\tau}_{-\nu}, \tau]$.
\end{inparaenum}
However, we also need the following lemmas in our proof..
\begin{lemma}
\label{epsi-tau} For every $\epsilon > 0$, there exists a
$\bar{\tau}_{\epsilon}$ such that for all $t \geq 0$, $0 \leq u \leq
\bar{\tau}_{\epsilon}$,
\[
\mathbb{P}\left(\Vert \bar{z}^K(t+u) - \bar{z}^K(t) \Vert >
\epsilon\right) = O(K^{-1}).
\]
\end{lemma}
\begin{IEEEproof}
Observe that
\begin{align*}
\lefteqn{\Vert \bar{z}^K(t+u)-\bar{z}^K(t) \Vert} \\
&~\leq \Vert \bar{z}^K(t)-\bar{z}(t) \Vert + \Vert \bar{z}^K(t+u)-\bar{z}(t+u) \Vert \\
&~ \ \ \ \ \ \ \ \ \ \ \ \ \ \ \ \ \ \ \ \ \ \ \ \ \ \ \ \ \ \ \ \ \ \ \ \ \ \ \ \ \ \ \ \ \ \ + \Vert \bar{z}(t)-\bar{z}(t+u) \Vert \\
&~\leq \Vert \bar{z}^K(t)-\bar{z}(t) \Vert + \Vert
\bar{z}^K(t+u)-\bar{z}(t+u) \Vert + \sqrt{2}\Lambda u
\end{align*}
Hence, for all $t \geq 0$, $u \geq 0$,
\begin{align*}
\lefteqn{\mathbb{P}\left(\Vert \bar{z}^K(t+u)-\bar{z}^K(t) \Vert >
\sqrt{2}\Lambda u + \frac{\epsilon}{2}\right)} \\
 &~\leq \mathbb{P}\left(\Vert \bar{z}^K(t)-\bar{z}(t) \Vert + \Vert
\bar{z}^K(t+u)-\bar{z}(t+u) \Vert > \frac{\epsilon}{2}\right) \\
&~\leq \mathbb{P}\left(\sup_{t \leq s \leq t+ u} \Vert
\bar{z}^K(s)-\bar{z}(s)
\Vert > \frac{\epsilon}{4} \right) \\
&~= O(K^{-1})
\end{align*}
where the last equality follows
from~\cite[Theorem~2.8]{stochproc.darling02fluid-limits}. Setting
$\bar{\tau}_{\epsilon} = \frac{\epsilon}{2\sqrt{2}\Lambda}$, for all
$t \geq 0$, $0 \leq u \leq \bar{\tau}_{\epsilon}$
\begin{align*}
\lefteqn{\mathbb{P}\left(\Vert \bar{z}^K(t+u)-\bar{z}^K(t) \Vert >
\epsilon
\right)} \\
&~\leq \mathbb{P}\left(\Vert \bar{z}^K(t+u)-\bar{z}^K(t) \Vert >
\sqrt{2}\Lambda u + \frac{\epsilon}{2} \right) \\
&~=O(K^{-1}).
\end{align*}
\end{IEEEproof}

\begin{lemma}
\label{epsi-delta} Suppose $u$ is a fixed time and $u^K$ is a random
time that satisfies $\mathbb{P}\left(|u- u^K| > \delta\right) =
O(K^{-1})$ for every $\delta
> 0$. Then, for every $\epsilon > 0$,
\[
\mathbb{P}\left(\Vert \bar{z}^K(u)- \bar{z}^K(u \wedge u^K) \Vert
> \epsilon\right) = O(K^{-1})
\]
\end{lemma}
\begin{IEEEproof}
Fix a $\delta > 0$. Then,
\begin{align*}
\lefteqn{\mathbb{P}(\Vert \bar{z}^K(u)- \bar{z}^K(u \wedge u^K)
\Vert
> \epsilon)}\\
&=\mathbb{P}(u-u^K > \delta) \\
& \ \ \ \ \ \ \ \ \ \ \ \ \ \ \ \ \mathbb{P}\left(\Vert
\bar{z}^K(u)-
\bar{z}^K(u \wedge u^K) \Vert > \epsilon \big| u-u^K > \delta\right)\\
&~ \ \ +\mathbb{P}(u-u^K \leq \delta) \\
 & \ \ \ \ \ \ \ \ \ \ \ \ \ \ \ \ \mathbb{P}\left(\Vert \bar{z}^K(u)-
\bar{z}^K(u \wedge u^K) \Vert > \epsilon \big| u-u^K \leq \delta\right)\\
&\leq O(K^{-1})+ \mathbb{P}\left(\Vert \bar{z}^K(u)-\bar{z}^K(u
\wedge u^K) \Vert > \epsilon \big| u-u^K \leq \delta\right) \\
&\leq O(K^{-1})+ \mathbb{P}\left(\Vert \bar{z}^K(u)- \bar{z}^K(u -
\delta) \Vert > \epsilon \big| u-u^K \leq \delta\right)
\end{align*}
where the last inequality holds because $\bar{z}^K(t)$ is a monotone
increasing function. Setting $\delta = \bar{\tau}_{\epsilon}$~(see
Lemma~\ref{epsi-tau}),
\begin{align*}
\lefteqn{\mathbb{P}(\Vert \bar{z}^K(u)- \bar{z}^K(u \wedge u^K)
\Vert
> \epsilon)}\\
&\leq O(K^{-1})+ \mathbb{P}\left(\Vert \bar{z}^K(u)-
\bar{z}^K(u - \bar{\tau}_{\epsilon}) \Vert > \epsilon \big| u-u^K \leq \bar{\tau}_{\epsilon}\right)\\
&= O(K^{-1})+ \mathbb{P}\left(\Vert \bar{z}^K(u)-
\bar{z}^K(u - \bar{\tau}_{\epsilon}) \Vert > \epsilon \right)\\
&\leq O(K^{-1}) + O(K^{-1}) \\
&= O(K^{-1})
\end{align*}
where the last inequality follows from Lemma~\ref{epsi-tau}.
\end{IEEEproof}

Following is the proof of Theorem~\ref{assym-optimality}.

\noindent
\begin{inparaenum}[(a)]
\item First, we prove the convergence of $z^K(t)$ to $z(t)$ over
$[0,\tau_{\nu}]$. Fix a $\nu > 0$.
Then Corollary~\ref{cor:apprx-bdry} implies that
$z^K(t)$ converges to $z(t)$ in the region $\mathcal{S}_{\nu}$. 
Following~\cite[Theorem~2.8]{stochproc.darling02fluid-limits} we
have, for all $\epsilon,\delta > 0$,
\begin{subequations}
\begin{align}
& \mathbb{P}\left(\sup_{0 \leq t \leq \tau_{\nu}} \Vert z^K(t
\wedge \tau^K_{\nu}) - z(t) \Vert > \epsilon \right) =
O(K^{-1}) \label{eqn:darling-a1}\\
&\mbox{and } \mathbb{P}(|\tau^K_{\nu} - \tau_{\nu}| > \delta) =
O(K^{-1}) \label{eqn:darling-a2} .
\end{align}
\end{subequations}
Since, for all $t \geq 0$,
\[
\Vert z^K(t)- z(t) \Vert \leq \Vert z^K(t \wedge \tau^K_{\nu}) -
z(t) \Vert + \Vert z^K(t) -  z^K(t \wedge \tau^K_{\nu}) \Vert,
\]
we obtain
\begin{align*}
\sup_{0 \leq t \leq \tau_{\nu}} \Vert z^K(t)- z(t) \Vert \leq
&~\sup_{0 \leq t \leq \tau_{\nu}} \Vert z^K(t \wedge \tau^K_{\nu}) -
z(t) \Vert \\
&~+ \sup_{0 \leq t \leq \tau_{\nu}} \Vert z^K(t) - z^K(t \wedge
\tau^K_{\nu}) \Vert.
\end{align*}
If the left side is larger than $\epsilon$, at least one of the two
terms on the right side is larger than $\epsilon/2$, and so by the
union bound, we get
\begin{align}
\lefteqn{\mathbb{P}\left(\sup_{0 \leq t \leq \tau_{\nu}} \Vert
z^K(t)- z(t) \Vert > \epsilon \right)} \nonumber \\
& \leq \mathbb{P}\left( \sup_{0 \leq t \leq \tau_{\nu}} \Vert z^K(t
\wedge \tau^K_{\nu}) - z(t) \Vert > \frac{\epsilon}{2}\right) \nonumber \\
&~ \ \ + \mathbb{P}\left( \sup_{0 \leq t \leq
\tau_{\nu}} \Vert z^K(t) - z^K(t \wedge \tau^K_{\nu}) \Vert > \frac{\epsilon}{2} \right) \nonumber\\
& \leq O(K^{-1}) + \mathbb{P}\left(\Vert z^K(\tau_{\nu}) -
z^K(\tau_{\nu}
 \wedge \tau^K_{\nu}) \Vert > \frac{\epsilon}{2} \right) \label{eqn:bound-1}
\end{align}
where the first term in the last inequality follows
from~\eqref{eqn:darling-a1}. Also, from
corollary~\ref{cor:apprx-bdry}, for $K \geq K_{\nu}$,
$\phi^K(z^K(\tau^K_{\nu})-) > 0$, i.e., the process $z^K(t)$ follows
uncontrolled dynamics until $\tau^K_{\nu}$. Thus, for $K \geq
K_{\nu}$, $z^K(\tau^K_{\nu}) = \bar{z}^K(\tau^K_{\nu})$ and
\[\Vert z^K(\tau_{\nu}) - z^K(\tau_{\nu}
 \wedge \tau^K_{\nu}) \Vert \leq \Vert \bar{z}^K(\tau_{\nu}) - \bar{z}^K(\tau_{\nu}
 \wedge \tau^K_{\nu}) \Vert \]
sample path wise. The inequality is an equality if $\tau_{\nu} \leq
\tau_{\nu}^K$;  both sides equal $0$ in this case. Otherwise, it is
an inequality because the possible change in dynamics of $z^K(t)$
after $\tau^K_{\nu}$ makes it increase (in both its components) at a
slower pace than the uncontrolled $\bar{z}^K(t)$. Thus
\begin{align*}
\lefteqn{\mathbb{P}\left(\Vert z^K(\tau_{\nu}) - z^K(\tau_{\nu}
 \wedge \tau^K_{\nu}) \Vert > \frac{\epsilon}{2} \right)} \\
&~\leq \mathbb{P}\left(\Vert \bar{z}^K(\tau_{\nu}) -
\bar{z}^K(\tau_{\nu}
 \wedge \tau^K_{\nu}) \Vert > \frac{\epsilon}{2} \right) \\
&~ \leq O(K^{-1})
\end{align*}
where the last inequality follows from~\eqref{eqn:darling-a2} and
Lemma~\ref{epsi-delta}. Using this in~\eqref{eqn:bound-1} we get
\begin{align*}
\mathbb{P}\left(\sup_{0 \leq t \leq \tau_{\nu}} \Vert z^K(t)- z(t)
\Vert > \epsilon \right) &~\leq O(K^{-1}) + O(K^{-1}) \\
&~ = O(K^{-1})
\end{align*}

\item Now we prove the convergence of $z^K(t)$ to $z(t)$ over
$[\tau_{\nu}, \tilde{\tau}_{-\nu}]$.
Observe that, for $t \in [\tau_{\nu}, \tilde{\tau}_{-\nu}]$,
\begin{align*}
\lefteqn{\Vert z^K(t)- z(t) \Vert } \\
&\leq \Vert z^K(\tau_{\nu})- z(\tau_{\nu}) \Vert +  \Vert z^K(t)- z^K(\tau_{\nu}) \Vert
 + \Vert z(t)- z(\tau_{\nu}) \Vert.
\end{align*}
Hence,
\begin{align*}
\lefteqn{\sup_{\tau_{\nu} \leq t \leq \tilde{\tau}_{-\nu}} \Vert z^K(t)- z(t) \Vert} \\
&~\leq \Vert z^K(\tau_{\nu})- z(\tau_{\nu}) \Vert + \sup_{\tau_{\nu} \leq t \leq \tilde{\tau}_{-\nu}} \Vert z^K(t)- z^K(\tau_{\nu}) \Vert \\
&~ \ \ \ \ \ \ \ \ \ \ \ \ \ \ \ \ \ \  \ \ \ \ \ \ \ + \sup_{\tau_{\nu} \leq t \leq \tilde{\tau}_{-\nu}} \Vert z(t)- z(\tau_{\nu}) \Vert \\
&~= \Vert z^K(\tau_{\nu})- z(\tau_{\nu}) \Vert + \Vert
z^K(\tilde{\tau}_{-\nu})- z^K(\tau_{\nu}) \Vert \\
&~ \ \ \ \ \ \ \ \ \ \ \ \ \ \ \ \ \ \ \ \ \  \ \ \ \ +  \Vert
z(\tilde{\tau}_{-\nu})- z(\tau_{\nu}) \Vert \\
&~\leq \Vert z^K(\tau_{\nu})- z(\tau_{\nu}) \Vert + \Vert
z^K(\tilde{\tau}_{-\nu})- z^K(\tau_{\nu}) \Vert + \sqrt{2}\Lambda
b\nu
\end{align*}
where the equality follows because the $z^(t)$ and $z(t)$ are
nondecreasing. The last inequality holds because $\Vert{\rm
d}{z}/{\rm d}t\Vert \leq \Vert{\rm d}{\bar{z}}/{\rm d}t\Vert \leq
\sqrt{2}\Lambda$ and $\tilde{\tau}_{-\nu} - \tau_{\nu} \leq b \nu$.
Moreover,
\begin{align*}
\mathbb{P}& \left(\sup_{\tau_{\nu} \leq t \leq \tilde{\tau}_{-\nu}}
 \Vert z^K(t)- z(t) \Vert > \sqrt{2}\Lambda
b\nu + \frac{\epsilon}{2} \right) \\
&~\leq \mathbb{P}\left(\Vert z^K(\tau_{\nu})- z(\tau_{\nu}) \Vert >
\frac{\epsilon}{4} \right) \\
&~ \ \ \ \  \ \ \ \  \ \ \ \ \ \ \ \ \ \ \ \ + \mathbb{P}\left(\Vert
z^K(\tilde{\tau}_{-\nu})- z^K(\tau_{\nu}) \Vert > \frac{\epsilon}{4}
\right) \\
&~=O(K^{-1}) + \mathbb{P}\left(\Vert z^K(\tilde{\tau}_{-\nu})-
z^K(\tau_{\nu}) \Vert > \frac{\epsilon}{4} \right)
\end{align*}
where the equality follows from the result of part~(a). We
now redefine the Markov process $\bar{z}^K(t) =
(\bar{x}^K(t),\bar{y}^K(t))$ for $t \geq \tau_{\nu}$, to be the
uncontrolled dynamics with initial condition $\bar{z}^K(\tau_{\nu})
= z^K(\tau_{\nu})$. Again, it can be easily observed that
\[\Vert z^K(\tilde{\tau}_{-\nu})-
z^K(\tau_{\nu}) \Vert \leq \Vert \bar{z}^K(\tilde{\tau}_{-\nu})-
\bar{z}^K(\tau_{\nu}) \Vert. \]
 Thus
\begin{align*}
\mathbb{P}& \left(\sup_{\tau_{\nu} \leq t \leq \tilde{\tau}_{-\nu}}
 \Vert z^K(t)- z(t) \Vert > \sqrt{2}\Lambda
b\nu + \frac{\epsilon}{2} \right) \\
&~\leq O(K^{-1}) + \mathbb{P}\left(\Vert \bar{z}^K(\tilde{\tau}_{-\nu})-
\bar{z}^K(\tau_{\nu}) \Vert > \frac{\epsilon}{4} \right)\\
&~ \leq O(K^{-1}) + \mathbb{P}\left(\Vert
\bar{z}^K(\tau_{\nu} + b\nu)- \bar{z}^K(\tau_{\nu}) \Vert >
\frac{\epsilon}{4} \right)
\end{align*}
Set $\nu = \min\{\frac{\epsilon}{2\sqrt{2}\Lambda
b},\bar{\tau}_{\frac{\epsilon}{4b}}\}$, and apply Lemma~\ref{epsi-tau} to get
\begin{align*}
\lefteqn{\mathbb{P} \left(\sup_{\tau_{\nu} \leq t \leq
\tilde{\tau}_{-\nu}}
 \Vert z^K(t)- z(t) \Vert > \epsilon \right)} \\
&~ \leq \mathbb{P}\left(\sup_{\tau_{\nu} \leq t \leq
\tilde{\tau}_{-\nu}}
 \Vert z^K(t)- z(t) \Vert > \sqrt{2}\Lambda
b\nu + \frac{\epsilon}{2} \right) \\
&~ \leq O(K^{-1}) + O(K^{-1}) \\
&~=O(K^{-1}).
\end{align*}

\item Finally, we prove the convergence of $z^K(t)$ to $z(t)$ over
$[\tilde{\tau}_{-\nu},\tau]$. Reconsider the process
$\tilde{z}^K(t), t \geq \tau_{\nu}$ and the associated function
$\tilde{z}(t)$. Recall that, for any $\nu > 0$, $\tilde{z}^K(t)$ and
$\tilde{z}(t)$ exit $\mathcal{S}_{-\nu}$ at $\tilde{\tau}^K_{-\nu}$
and $\tilde{\tau}_{-\nu}$ respectively. Clearly,
$\tilde{\tau}_{-\nu/2} < \tilde{\tau}_{-\nu}$; say
$\tilde{\tau}_{-\nu} - \tilde{\tau}_{-\nu/2} = \delta_{\nu}$. Also,
using~\cite[Theorem~2.8]{stochproc.darling02fluid-limits},
\begin{align*}
&~ \mathbb{P}\left(\tilde{\tau}^K_{-\nu/2}- \tilde{\tau}_{-\nu/2} > \delta_{\nu}\right) = O(K^{-1})\\
\mbox{i.e., } &~ \mathbb{P}\left(\tilde{\tau}^K_{-\nu/2} >  \tilde{\tau}_{-\nu} \right) = O(K^{-1})
\end{align*}
Furthermore, we have that $\tau^K_{-\nu/2} \leq
\tilde{\tau}^K_{-\nu/2}$ sample path wise. The  inequality holds
because $z^K(t)$ may continue to increase (in both its components)
at a higher pace than $\tilde{z}^K(t)$ even after $\tau_{\nu}$. Thus
\[\mathbb{P}\left(\tau^K_{-\nu/2} > \tilde{\tau}_{-\nu} \right) = O(K^{-1}),\]
implying that the probability that  $z^K(t)$ has changed its
dynamics by $\tilde{\tau}_{-\nu}$ approaches $1$ as $K$ approaches
$\infty$. In these realizations, the dynamics of $z^K(t)$ and $z(t)$
match for $t \geq \tilde{\tau}_{-\nu}$. We restrict ourselves to
only these realizations. We also have from part~(b) that, for every
$\epsilon
> 0$,
\[
\mathbb{P}\left(\Vert z^K(\tilde{\tau}_{-\nu})- z(\tilde{\tau}_{-\nu}) \Vert > \epsilon \right) = O(K^{-1})
\]
Once more using~\cite[Theorem~2.8]{stochproc.darling02fluid-limits}, for any $\epsilon,\delta > 0$
\begin{align*}
& \mathbb{P}\left(\sup_{\tilde{\tau}_{-\nu} \leq t \leq \tau} \Vert
z^K(t)- z(t) \Vert > \epsilon \right) = O(K^{-1}) \\
\mbox{and } & \mathbb{P}\left(|\tau^K - \tau| > \delta \right) = O(K^{-1}).
\end{align*}

\end{inparaenum}

\section{Proof of Theorem~\ref{theorem:asym-optimal}}
\label{proof-asym-optimal}
For the optimal policy $u^{\ast}$, the total expected cost
\begin{equation*}
\mathbb{E}^K_{u^{\ast}}\{\mathcal{T}_d + \gamma\mathcal{E}_c\} = \mathbb{E}^K_{u^{\ast}}\{\tau^K + \Gamma (X + y^K(\tau^K))\}
\end{equation*}
since $\mathcal{T}_d = \tau^K$ by definition~(see~\eqref{eqn:tau-K});
we use the subscript $u^{\ast}$ to show dependence of the probability law on the underlying policy.
Under the deterministic policy $u^{\infty}$, copying to relays is stopped at the deterministic
time instant $\tau^{\ast} < \tau$, implying $y^K(\tau^{\ast})=y^K(\tau)$. Thus, the total expected cost
\begin{equation*}
\mathbb{E}^K_{u^{\infty}}\{\mathcal{T}_d + \gamma\mathcal{E}_c\} =
\mathbb{E}^K_{u^{\infty}}\{\tau^K + \Gamma (X + y^K(\tau))\}.
\end{equation*}
Also observe that for $(x^K(t),y^K(t))$ under $u^{\infty}$, the corresponding fluid limits are
the same deterministic dynamics $(x(t),y(t))$ defined in
Section~\ref{asym-det-dynamics}~(i.e., solutions of~\eqref{eqn:f-1}-\eqref{eqn:f-2}).
$(x^K(t),y^K(t))$ and $(x(t),y(t))$ satisfy the hypotheses assumed in Darling~\cite{stochproc.darling02fluid-limits} over the intervals
$[0,\tau^{\ast}]$ and $[\tau^{\ast},\infty)$. Thus~\cite[Theorem~2.8]{stochproc.darling02fluid-limits}
applies, and we conclude~\footnote{Applying~\cite[Theorem~2.8]{stochproc.darling02fluid-limits} over $[0,\tau^{\ast}]$ yields
$\lim_{K \rightarrow \infty} \mathbb{P}\left(\Vert(x^K(\tau^{\ast}),y^K(\tau^{\ast})
- (x(\tau^{\ast}),y(\tau^{\ast}))\Vert > \epsilon \right) = 0$
which is a necessary condition to apply~\cite[Theorem~2.8]{stochproc.darling02fluid-limits} over $[\tau^{\ast},\infty)$.}
\begin{align*}
&\lim_{K \rightarrow \infty} \mathbb{P}^K_{u^{\infty}}\left(\sup_{0 \leq t \leq \tau}\Vert(x^K(t),y^K(t)) - (x(t),y(t))\Vert > \epsilon \right) = 0, \\
&\lim_{K \rightarrow \infty} \mathbb{P}^K_{u^{\infty}}\left(|\tau^K
- \tau| > \delta \right) = 0.
\end{align*}

\remove{
Furthermore, it can be easily shown that under both the controls $u^{\ast}$ and $u^{\infty}$,
the delivery delays $\tau^K$ have second moments which are bounded uniformly over all $K$, i.e.,\footnote{The proof entails
binding $\tau^K$~(under the control $u^{\ast}$ or $u^{\infty}$) by the delivery delays under the policy that never
copies to relays, and showing that the latter have second moments which are bounded uniformly over all $K$.}
\begin{align*}
\sup_K\mathbb{E}^K_{u^{\ast}}(\tau^K)^2 < \infty,\\
\sup_K\mathbb{E}^K_{u^{\infty}}(\tau^K)^2 < \infty.
\end{align*}
}
Furthermore, it can be easily shown that under both the controls $u^{\ast}$ and $u^{\infty}$,
the delivery delays $\tau^K$ have second moments that are bounded uniformly over all $K$. To see this,
consider a policy $u^0$ that never copies to relays. Clearly,
\begin{align*}
\mathbb{E}^K_{u^{\ast}}(\tau^K)^2 < &\mathbb{E}^K_{u^0}(\tau^K)^2,\\
\mathbb{E}^K_{u^{\infty}}(\tau^K)^2 < &\mathbb{E}^K_{u^0}(\tau^K)^2
\end{align*}
for each $K$. Then is suffices to show that
\begin{equation}
\sup_K\mathbb{E}^K_{u^0}(\tau^K)^2 < \infty. \label{eqn:delay-u0}
\end{equation}
Note that
\begin{equation*}
\tau^K = \sum_{m = 0}^{M^K_{\alpha}-1}\bar{\delta}_m
\end{equation*}
where $\bar{\delta}_m$ is the time duration for which $m(t) = m$; $\bar{\delta}_m, m = 0,1,\dots$ are independent,
and $\bar{\delta}_m$  is exponentially distributed with mean
$\frac{1}{\lambda^K(m + N^K_0)(M^K - m)}$ under policy  $u^0$.
Thus
\begin{align*}
\mathbb{E}^K_{u^0}\tau^K = &\sum_{m = 0}^{M^K_{\alpha}-1}\frac{1}{\lambda^K(m + N^K_0)(M^K - m)} \\
                         \leq &\sum_{m = 0}^{M^K_{\alpha}-1}\frac{1}{\lambda^K N^K_0(M^K - M^K_{\alpha})} \\
                         = & \frac{M^K_{\alpha}}{\lambda^K N^K_0 (M^K - M^K_{\alpha})} \\
                         = &\frac{X_{\alpha}}{\Lambda Y_0(X - X_{\alpha})} \\
                         < &\infty
\end{align*}
Similarly,
\begin{align*}
\mbox{Var}^K_{u^0}\tau^K = &\sum_{m = 0}^{M^K_{\alpha}-1}\frac{1}{\left(\lambda^K(m + N^K_0)(M^K - m)\right)^2} \\
                        \leq & \frac{M^K_{\alpha}}{\left(\lambda^K N^K_0 (M^K - M^K_{\alpha})\right)^2} \\
                         = &\frac{X_{\alpha}}{K\Lambda^2 Y^2_0(X - X_{\alpha})^2} \\
                         \rightarrow & 0
\end{align*}
as $K \rightarrow \infty$. These results together imply~\eqref{eqn:delay-u0}.

Following~\cite[Remark~9.5.1]{stochproc.measure-and-probability}, under both $u^{\ast}$ and $u^{\infty}$,
$\tau^K$ are uniformly integrable. Since, $\tau^K$, under both $u^{\ast}$ and $u^{\infty}$, converge to $\tau$
in probability and hence in distribution,~\cite[Theorem~9.5.1]{stochproc.measure-and-probability} yields
\begin{equation}
\lim_{K \rightarrow \infty} \mathbb{E}^K_{u^{\ast}} \tau^K =  \lim_{K \rightarrow \infty} \mathbb{E}^K_{u^{\infty}} \tau^K  = \tau \label{eqn:delay-tau}.
\end{equation}

Next, it is easy to show that under the control $u^{\ast}$, $y^K(\tau^K)$ converges
to $y(\tau)$ in probability. To see this, observe that
\begin{equation}
|y^K(\tau^K) - y(\tau)| \leq |y^K(\tau^K) - y^K(\tau)| + | y^K(\tau) - y(\tau)|. \label{ineq:conv-y}
\end{equation}
From Theorem~\ref{assym-optimality}, $y^K(\tau)$ and $\tau^K$ converge to $y(\tau)$ and $\tau$ respectively, in probability.
The latter result, along with the arguments similar to those in the proof of Lemma~\ref{epsi-delta}, implies
that
\[
\mathbb{P}\left(| y^K(\tau^K) - y^K(\tau) | > \epsilon\right) = O(K^{-1})
\]
for every $\epsilon > 0$. Using these facts in~\eqref{ineq:conv-y}, we conclude that
\[
\mathbb{P}\left(| y^K(\tau^K) - y(\tau) | > \epsilon\right) = O(K^{-1}).
\]
for every $\epsilon > 0$. Since $y^K(\tau^K)$ is bounded, and hence uniformly integrable,~\cite[Theorem~9.5.1]{stochproc.measure-and-probability}
implies that
\begin{equation}
\lim_{K \rightarrow \infty} \mathbb{E}^K_{u^{\ast}} y^K(\tau^K) = y(\tau). \label{eqn:conv-y_ast}
\end{equation}
Similarly, under the control $u^{\infty}$ also, $y^K(\tau)$ is bounded, and hence is uniformly integrable. It
also converges to  $y(\tau)$ in probability. Once more using~\cite[Theorem~9.5.1]{stochproc.measure-and-probability}, we get
\begin{equation}
\lim_{K \rightarrow \infty} \mathbb{E}^K_{u^{\infty}} y^K(\tau) = y(\tau). \label{eqn:conv-y_infty}
\end{equation}
Combining~\eqref{eqn:conv-y_ast} and~\eqref{eqn:conv-y_infty}
\begin{equation}
\lim_{K \rightarrow \infty} \mathbb{E}^K_{u^{\ast}} y^K(\tau^K) =
\lim_{K \rightarrow \infty} \mathbb{E}^K_{u^{\infty}} y^K(\tau).
\label{eqn:copy-cost}
\end{equation}
Finally, combining~\eqref{eqn:delay-tau} and~\eqref{eqn:copy-cost}, we get that
\[
\lim_{K \rightarrow \infty} \mathbb{E}^K_{u^{\ast}}\{\mathcal{T}_d + \gamma\mathcal{E}_c\} =  \lim_{K \rightarrow \infty} \mathbb{E}^K_{u^{\infty}}\{\mathcal{T}_d + \gamma\mathcal{E}_c\} = \tau + \Gamma y(\tau^{\ast}).
\]

\section{The Hamiltonian Formulation and The Solution}
\label{hamiltonian}

In this section we consider the limiting deterministic~(fluid) system  
and study its optimal control.
The limiting controlled system is: $x(0) = 0$, $y(0) = Y_0$, and for $t \geq 0$,
\begin{subequations}
\begin{align}
\frac{{\rm d}x(t)}{{\rm d}t} &= \Lambda (x(t) + y(t))(X-x(t)), \label{controlled-drift-1}\\
\frac{{\rm d}y(t)}{{\rm d}t} &=  \Lambda (x(t) + y(t))(Y-y(t))u(t) \label{controlled-drift-2}\
\end{align}
\end{subequations}
where $u(t) \in [0,1]$ is the control at time $t$. Our objective is to minimize
\begin{equation}
\Gamma y(T) + T = \Gamma y(T) + \int_0^{T}1 . {\rm d}t \label{objective}
\end{equation}
where $T$ is the terminal time when $x(T) = X_{\alpha}$; dependence of $T$ on the underlying control is understood, and
is not shown explicitly. 
\begin{theorem}
The optimal policy for the deterministic system~\eqref{controlled-drift-1}-\eqref{controlled-drift-2} with cost~\eqref{objective} 
is 
\[u^{\ast}(t) = 1_{[0,\tau^{\ast}]}(t)\]
with $\tau^{\ast}$ as in~\eqref{eqn:stop-relays}.
Furthermore, the optimal cost is $\tau + \Gamma y(\tau^{\ast})$ with  $\tau$ as in~\eqref{eqn:stop-dstns}.
\end{theorem}
\begin{IEEEproof}
Following~\cite[Section~3.3.1]{stochctrl.bertsekas05dpoc-vol1}, we define the Hamiltonian for the system
\begin{align}
\lefteqn{H(x,y,u,p_1,p_2)} \nonumber \\
 &= 1 + p_1\Lambda(X-x)(x+y) + p_2 \Lambda(Y-y)(x+y)u \nonumber \\
 &= 1 + \Lambda(x+y)[p_1(X-x) + p_2(Y-y)u] \label{eqn:hamiltonian}
\end{align}
where $p_i:\mathbb{R}_+ \rightarrow \mathbb{R}, i = 1,2$ are the cojoint functions
associated with $x(t)$ and $y(t)$ respectively.
Let $u^{\ast}(t), t \geq 0$, be an optimal control trajectory.
Let $T^{\ast}$ be the corresponding terminal time, and let $(x^{\ast}(t),y^{\ast}(t)), t \in [0,T^{\ast}]$
be the  corresponding state trajectory.
\paragraph{Adjoint equations}
By~\cite[Section~3.3.1, Proposition~3.1]{stochctrl.bertsekas05dpoc-vol1},
the functions $p_i(t)$ are
solutions of the following adjoint equations:
\begin{align}
\lefteqn{\frac{{\rm d}p_1(t)}{{\rm d}t}=-\left.\frac{\partial}{\partial x}H(x,y^{\ast},u^{\ast},p_1,p_2)\right|_{x = x^{\ast}}=}\nonumber\\
 &-\Lambda[p_1(t)(X-2x^{\ast}(t) - y^{\ast}(t)) + p_2(t)(Y-y^{\ast}(t))u^{\ast}(t)], \label{adjoint1} \\
\lefteqn{\frac{{\rm d}p_2(t)}{{\rm d}t}=-\left.\frac{\partial}{\partial y}H(x^{\ast},y,u^{\ast},p_1,p_2)\right|_{y = y^{\ast}}=}\nonumber\\
 &-\Lambda[p_1(t)(X-x^{\ast}(t)) + p_2(t)(Y - x^{\ast}(t) -2 y^{\ast}(t))u^{\ast}(t)]. \label{adjoint2}
\end{align}
\paragraph{Boundary condition} 
Observe that the terminal cost is $\Gamma y^{\ast}(T^{\ast})$.
Thus, by~\cite[Section~3.3.1, Proposition~3.1]{stochctrl.bertsekas05dpoc-vol1},
\begin{equation}
p_2(T^{\ast}) = \left. \frac{\partial}{\partial y} \left(\Gamma
y\right) \right|_{y = y^{\ast}(T^{\ast})} = \Gamma.
\label{adjoint-bdary-cond1}
\end{equation}
\paragraph{Minimum principle} Moreover, the optimal control $u^{\ast}$ satisfies
\begin{equation*}
u^{\ast}(t) = \arg \min_{u \in [0,1]} H(x^{\ast}(t),y^{\ast}(t),u,p_1(t),p_2(t)) 
 \end{equation*}
for all $t \in [0,T^{\ast}]$. From~\eqref{eqn:hamiltonian}, it is
immediate that the optimal policy is a bang-bang policy.
\begin{align}
u^{\ast}(t) = \left\{ \begin{array}{ll}
                 1, \mbox{ if } p_2(t) \leq 0 \\
                 0, \mbox{ if } p_2(t) > 0\end{array} \right. \label{hamiltonian-min}
\end{align}
In particular, our observation~\eqref{adjoint-bdary-cond1} implies that
$u^{\ast}(T^{\ast}) = 0$. 
\paragraph{Free terminal time condition}
Since the terminal time is free, we also
have from~\cite[Section~3.4.3]{stochctrl.bertsekas05dpoc-vol1} that
\[
H(x^{\ast}(t),y^{\ast}(t),u^{\ast}(t),p_1(t),p_2(t)) = 0
\]
for all $t \in [0, T^{\ast}]$. In particular, equality at $t =
T^{\ast}$ implies~(see~\eqref{eqn:hamiltonian})
\[
1 + \Lambda(X_{\alpha} + y(T^{\ast}))[p_1(T^{\ast})(X - X_{\alpha})]
= 0.
\]
Since $X - X_{\alpha} > 0$, we must have
\begin{equation}
p_1(T^{\ast}) < 0. \label{adjoint-bdary-cond2}
\end{equation}
 We will find this observation useful later.

Our characterization of the optimal control consists of two steps.
First we show that the optimal control trajectory is of threshold
type, i.e.,
\begin{align}
u^{\ast}(t) = \left\{ \begin{array}{ll}
                 1, \mbox{ if } t \in [0, t^{\ast}]\\
                 0, \mbox{ if } t \in (t^{\ast},T^{\ast}].\end{array}
                 \right. \label{threshold-policy}
\end{align}
This is done in the next subsection. In the subsequent subsection,
we obtain the threshold $t^{\ast}$.

\subsection{Optimal control is of threshold type}
We show that $p_2(t)$ is negative for $t \in [0, t^{\ast}]$ and
strictly positive for $t \in (t^{\ast}, T^{\ast}]$ for some
$t^{\ast} \geq 0$. It then follows from~\eqref{hamiltonian-min} that
$u^{\ast}(t)$ is as in~\eqref{threshold-policy}. Recall $\frac{{\rm
d}p_1(t)}{{\rm d}t}$ in~\eqref{adjoint1}. We consider two scenarios.
\subsubsection{Case~1}
\label{form-part1} Let $X - 2X_{\alpha} - y^{\ast}(T^{\ast}) \geq
0$. Since $x^{\ast}(t)$ and $y^{\ast}(t)$ both are non-decreasing in
$t$, we have
\[
X - 2x^{\ast}(t) - y^{\ast}(t) \geq 0 \mbox{ for all } t \in [0,T^{\ast}].
\] 
Moreover, from~\eqref{hamiltonian-min},
\[
p_2(t)u^{\ast}(t) \leq 0 \mbox{ for all } t \in [0,T^{\ast}]
\]
with equality at $t = T^{\ast}$. Thus, from~\eqref{adjoint1},
\[
\frac{{\rm d}p_1(t)}{{\rm d}t} \geq 0 
\]
for all  $t \in [t',T^{\ast}]$ at which $p_1(t) < 0$. But, using the observation
$p_1(T^{\ast}) < 0$~(see~\eqref{adjoint-bdary-cond2}), it immediately follows that
\[
\frac{{\rm d}p_1(t)}{{\rm d}t} \geq 0 \mbox{ for all } t \in [0,T^{\ast}],
\]
and so, $p_1(t) < 0$ for all $t \in [0,T^{\ast}]$. Now, from~\eqref{adjoint2}, 
\[
\frac{{\rm d}p_2(t)}{{\rm d}t} > 0 
\]
for all  $t \in [0,T^{\ast}]$ at which $p_2(t) \geq 0$.  Again, using the observation
$p_2(T^{\ast}) = \Gamma > 0$~(see~\eqref{adjoint-bdary-cond1}), it follows that
either $p_2(t) > 0$ for all $t \in [0,T^{\ast}]$, 
or there exists a $t^{\ast} \in [0, T^{\ast}]$ such that $p_2(t^{\ast}) = 0$, and 
\begin{align*}
p_2(t) \left\{ \begin{array}{ll}
                 < 0, \mbox{ if } t \in [0, t^{\ast})\\
                 > 0, \mbox{ if } t \in (t^{\ast},T^{\ast}].\end{array} \right.
\end{align*}

\subsubsection{Case 2}
\label{form-part2}
 Let $X - 2X_{\alpha} - y^{\ast}(T^{\ast}) < 0$.
Observe that $X - 2x^{\ast}(t) - y^{\ast}(t)$ is decreasing in $t$.
Thus, tracing back from $t = T^{\ast}$, there exists a $t_1$ such
that $X - 2x^{\ast}(t_1) - y^{\ast}(t_1) = 0$; we set $t_1 = 0$ if
$X - 2x^{\ast}(t) - y^{\ast}(t) < 0$ for all $t \in [0, T^{\ast}]$.
Clearly, $X - 2x^{\ast}(t) - y^{\ast}(t) \leq 0$ for all $t \in
[t_1, T^{\ast}]$.

We claim that $p_1(t) < 0$  for all $t \in [t_1, T^{\ast}]$. 
Suppose not, i.e., there exists a $t_2 \in [t_1,T^{\ast}]$ such that
$p_1(t_2) \geq 0$. Then, from~\eqref{adjoint1},
\[ 
\frac{{\rm d}p_1(t)}{{\rm d}t} \geq 0 \mbox{ for all } t \in [t_2,T^{\ast}],
\]
and so, $p_1(t)$ increases with $t$ in this interval. But this
contradicts the assertion in~\eqref{adjoint-bdary-cond2} that
$p_1(T^{\ast}) < 0$. Hence the claim holds.

Now, $X - 2x^{\ast}(t_1) - y^{\ast}(t_1) =0$, and $p_1(t_1) < 0$.
An argument similar to that in {\it Case~1} yields that 
\[
\frac{{\rm d}p_1(t)}{{\rm d}t} \geq 0 \mbox{ for all } t \in [0,t_1],
\]
and so, $p_1(t) < 0$ for all $t \in [0,T^{\ast}]$; 
recall that it is readily seen that $p_1(t) < 0$  for all $t \in [t_1, T^{\ast}]$.  
Consequently, as in {\it Case~1},
either $p_2(t) > 0$ for all $t \in [0,T^{\ast}]$, 
or there exists a $t^{\ast} \in [0, T^{\ast}]$ such that $p_2(t^{\ast}) = 0$, and 
\begin{align*}
p_2(t) \left\{ \begin{array}{ll}
                 < 0, \mbox{ if } t \in [0, t^{\ast})\\
                 > 0, \mbox{ if } t \in (t^{\ast},T^{\ast}].\end{array} \right.
\end{align*}

\remove{
$p_1(t) \leq p_1(t_1)$ for all $t \in [0, t_1]$. Our assertion
for $t $conclusion implies that~(see~\eqref{adjoint2})
\[
\frac{{\rm d}p_2(t)}{{\rm d}t} > 0 \mbox{ for all } t \in
[t_1,T^{\ast}],
\]
and so, $p_2(t)$ is increasing in $[t_1,T^{\ast}]$. Since
$p_2(T^{\ast}) = \Gamma >0$, either there exists a $t^{\ast} \in
[t_1, T^{\ast}]$ such that $p_2(t^{\ast}) = 0$ and
\begin{align*}
p_2(t) \left\{ \begin{array}{ll}
                 < 0, \mbox{ if } t \in [t_1,t^{\ast})\\
                 > 0, \mbox{ if } t \in (t^{\ast},T^{\ast}],\end{array} \right.
\end{align*}
or $p_2(t) > 0$ for all $t \in [t_1,T^{\ast}]$. In the former case,

\[
\frac{{\rm d}p_2(t)}{{\rm d}t} > 0.
\]
for all  $t \in [0,t_1]$, further implying that $p_2(t) < 0$ for all
$t \in [0,t_1]$. In the latter case, $X - 2x^{\ast}(t_1) -
y^{\ast}(t_1) =0 $. We also have $p_1(t_1) < 0$ and $p_2(t_1) > 0$.
But this is similar to {\it Case~1}, whence we conclude that either
$p_2(t) > 0$ for all $t \in [0,t_1]$, or there exists a $t^{\ast}
\in [0, t_1)$ such that
\begin{align*}
p_2(t) \left\{ \begin{array}{ll}
                 < 0, \mbox{ if } t \in [0, t^{\ast})\\
                 > 0, \mbox{ if } t \in (t^{\ast},t_1].\end{array} \right.
\end{align*}
}

To summarize, in both the cases there exits a $t^{\ast} \in
[0,T^{\ast}]$ such that
\begin{align*}
p_2(t) \left\{ \begin{array}{ll}
                 < 0, \mbox{ if } t \in [0, t^{\ast})\\
                 > 0, \mbox{ if } t \in (t^{\ast},T^{\ast}].\end{array} \right.
\end{align*}

\subsection{Optimum Threshold}
We now characterize the optimal threshold $t^{\ast}$. Consider a threshold policy
\begin{align*}
u(t) = \left\{ \begin{array}{ll}
                 1, \mbox{ if } t \in [0, \bar{t}~]\\
                 0, \mbox{ if } t \in (\bar{t},T].\end{array} \right.
\end{align*}
Let the corresponding state trajectory be
$(x^{\bar{t}}(t),y^{\bar{t}}(t)), t \geq 0$, and let the terminal
time be $T(\bar{t})$. Let $\bar{x} := x^{\bar{t}}(\bar{t})$ and
$\bar{y} := y^{\bar{t}}(\bar{t})$ be the values at the threshold
time $\bar{t}$. Clearly,
\begin{equation}
\frac{{\rm d}\bar{x}}{{\rm d}\bar{t}} = \Lambda(\bar{x} +
\bar{y})(X-\bar{x}). \label{eqn:gredient-barx}
\end{equation}
The associated cost is
\begin{equation}
C(\bar{t}) = T(\bar{t}) + \Gamma \bar{y}, \label{eqn:cost-threshold}
\end{equation}
and\footnote{We can restrict to only those $\bar{t}$ such that $\bar{x} := x^{\bar{t}}(\bar{t}) \leq X_{\alpha}$.}
\[
t^{\ast} = \arg \min_{\bar{t} \geq 0} C(\bar{t}).
\]
For any $\bar{t} \geq 0$ and $t \in (\bar{t}, \infty)$,
\begin{align*}
y^{\bar{t}}(t) &= \bar{y}, \\
\mbox{and } \frac{{\rm d}x^{\bar{t}}(t)}{{\rm d}t} &= \Lambda
(x^{\bar{t}}(t) + \bar{y})(X-x^{\bar{t}}(t)),
\end{align*}
and so
\[
T(\bar{t}) = \bar{t} + \frac{1}{\Lambda}\int_{\bar{x}}^{X_{\alpha}}\frac{{\rm d}z}{(z + \bar{y})(X - z)}.
\]
Its substitution in~\eqref{eqn:cost-threshold} yields
\[
C(\bar{t}) = \bar{t} +  \Gamma \bar{y} + \frac{1}{\Lambda}\int_{\bar{x}}^{X_{\alpha}}\frac{{\rm d}z}{(z + \bar{y})(X - z)}.
\]
Using Leibniz rule of differentiation, we get
\begin{align*}
\frac{{\rm d}C(\bar{t})}{{\rm d}\bar{t}} &= 1 + \Gamma \frac{{\rm d}\bar{y}}{{\rm d}\bar{t}} -\frac{1}{\Lambda} \left[\frac{{\rm d}\bar{y}}{{\rm d}\bar{t}}\int_{\bar{x}}^{X_{\alpha}}\frac{{\rm d}z}{(z + \bar{y})^2(X - z)} \right. \\
     & \ \ \ \ \ \ \ \ \ \ \ \ \ \ \ \ \ \ \ \ \ \ \left. + \frac{{\rm d}\bar{x}}{{\rm d}\bar{t}}\frac{1}{(\bar{x} + \bar{y})(X - \bar{x})} \right]      \\
 &=  \frac{{\rm d}\bar{y}}{{\rm d}\bar{t}} \left[\Gamma - \frac{1}{\Lambda}\int_{\bar{x}}^{X_{\alpha}}\frac{{\rm d}z}{(z + \bar{y})^2(X - z)} \right]
\end{align*}
where the last equality uses~\eqref{eqn:gredient-barx}. Defining
\[g(\bar{t}) := \Gamma - \frac{1}{\Lambda}\int_{\bar{x}}^{X_{\alpha}}\frac{{\rm d}z}{(z + \bar{y})^2(X - z)},\]
we get
\[\frac{{\rm d}C(\bar{t})}{{\rm d}\bar{t}} = \frac{{\rm d}\bar{y}}{{\rm d}\bar{t}}g(\bar{t}).\]
Note that $\frac{\partial g}{\partial \bar{x}} > \frac{1}{X - X_{\alpha}}$, $\frac{\partial g}{\partial \bar{y}} \geq 0$,
$\frac{{\rm d}\bar{x}}{{\rm d}\bar{t}} > \Lambda Y_0(X - X_{\alpha})$, $\frac{{\rm d}\bar{y}}{{\rm d}\bar{t}} \geq 0$, 
and so $g(\bar{t})$ is also strictly increasing in $\bar{t}$ with slope bounded away from $0$.
Thus, the optimal threshold is given by
\begin{align*}
t^{\ast} = \left\{ \begin{array}{ll}
                 0 &\mbox{if } g(0) > 0,\\
                 g^{-1}(0) &\mbox{otherwise}\end{array} \right.
\end{align*}
which is identical to $\tau^{\ast}$ in~\eqref{eqn:stop-relays}.
\end{IEEEproof}
\begin{remarks}
Combined with Theorem~\ref{theorem:asym-optimal}, we now have that the limit of the optimal cost~(of the finite problem) equals the
optimal cost of the limiting system. This does not hold in general~(see Remark~\ref{remark:cost-comparision}).
\end{remarks}

\bibliographystyle{IEEEtran}
\bibliography{IEEEabrv,ctrl-theory,stoch-proc,stoch-ctrl,comm-net}

\end{document}